\documentclass[12pt]{article}
\usepackage{amsfonts,graphicx,amsmath,amsthm,amssymb}
\usepackage{epsfig}
\usepackage{float}
\allowdisplaybreaks
\begin{document}

\title{\bf Applicability of Modified Gauss-Bonnet Gravity Models on the
Existence of Stellar Structures}
\author{K. Hassan \thanks{komalhassan3@gmail.com}~,
Tayyab Naseer \thanks{tayyabnaseer48@yahoo.com;
tayyab.naseer@math.uol.edu.pk}~
and M. Sharif \thanks{msharif.math@pu.edu.pk} \\
Department of Mathematics and Statistics, The University of Lahore,\\
1-KM Defence Road Lahore-54000, Pakistan.}

\date{}
\maketitle

\begin{abstract}
In this paper, we explore the existence of spherically symmetric
strange quark configurations coupled with anisotropic fluid setup in
the framework of modified Gauss-Bonnet theory. In this regard, we
adopt two models such as \emph{(i)}
$f(\mathcal{G})=\beta\mathcal{G}^2$, and \emph{(ii)}
$f(\mathcal{G})=\delta_{1}\mathcal{G}^{x}(\delta_{2}\mathcal{G}^{y}+1)$,
and derive the field equations representing a static sphere. We then
introduce bag constant in the gravitational equations through the
use of MIT bag model, so that the quarks' interior can be discussed.
Further, we work out the modified equations under the use of Tolman
IV ansatz to make their solution possible. Junction conditions are
also employed to find the constants involved in the considered
metric potentials. Afterwards, different values of model parameters
and bag constant are taken into account to graphically exploring the
resulting solutions. This analysis is done by considering five
strange quark objects like Her X-I, LMC X-4, 4U 1820-30, PSR J
1614-2230, and Vela X-I. Certain tests are also applied on the
developed models to check their physical feasibility. It is much
interesting that this modified gravity under its both considered
functional forms yield physically viable and stable results for
certain parametric values.
\end{abstract}

\section{Introduction}

The universe is home to a diverse array of celestial objects, from
the simplest to the most complex. The presence of extraordinary
elements throughout the cosmos fuels the curiosity of researchers,
who strive to unreveal the unique properties of these components in
hopes of shedding light on the enigmas of the universe. The core of
celestial bodies is the site of numerous nuclear reactions, which
generate the heat and illumination that permeate the cosmos. During
this, a celestial body reaches at a point where the inward pull of
gravity overcomes the outward force of pressure. This causes the
object to explode, leading to the formation of unique stellar
remnants. One of them is the neutron stars in which the intense
gravitational force compresses matter to the point where electrons
combine with protons, creating a sea of neutrons. These neutrons
exhibit quantum degeneracy, exerting an opposing force to gravity
that prevents the star from collapsing further. Neutron stars, which
are incredibly dense stellar objects, were theoretically predicted
in 1934, even though their actual existence was confirmed at a later
stage \cite{1}. Interestingly, between the realms of black holes and
neutron stars, there exists a hypothetical state known as a strange
star. These strange stars are believed to be even denser than
neutron stars, with their interiors composed of a unique combination
of different (strange, down, and up) quarks. Several researchers
have made attempts to understand and study these fictitious bodies
\cite{2}-\cite{2b}.

Studying anisotropic systems, where the physical characteristics
vary with direction, can provide valuable insights into the
structure and evolution of astronomical bodies. Numerous physical
phenomena occur that cause these systems to depart from uniform
isotropy, where the properties are the same in all directions
\cite{45}. Jeans \cite{46} introduced innovative concepts suggesting
that anisotropy could arise from diverse factors, implying
non-uniform pressure distribution within the interior. This
non-uniformity manifests when pressures acting tangentially and
radially differ, leading to the emergence of anisotropy. Ruderman
\cite{3} postulated the existence of massive interstellar entities
with densities exceeding $10^{15}$g/cm$^3$ and inferred that these
objects could exhibit anisotropic characteristics based on
theoretical analyzes. Research has shown that the presence of a
powerful magnetic field \cite{47}-\cite{47b} or other influential
factors \cite{48}-\cite{48b} around a star can lead to anisotropy.
Furthermore, Herrera's work \cite{49} demonstrated that the
emergence of dissipative flux, inhomogeneous energy density, or
shear, either individually or in combination, can disrupt the
isotropic structure within a star, resulting in anisotropy.

Anisotropic fluid distributions showcase a variety of captivating
physical processes, which are thought to be the essential
constituents of more realistic compact structures. These processes
encompass the presence of a superfluid, or boson stars
\cite{53,53a}, etc. In a study conducted by Dev and Gleiser
\cite{5}, specific equations of state (EoSs) were employed to
investigate the anisotropic behavior exhibited by celestial objects.
To investigate the characteristics of 4U 1820-30, Hossein and
colleagues \cite{6} incorporated $\Lambda$ into the gravitational
equations and generated graphical models to analyze the system's
properties. Kalam et al. \cite{7} evaluated the feasibility of
various neutron star configurations, assessing their potential to
exist as stable celestial objects. Maurya et al. \cite{8} employed
the embedding condition in order to determine the conditions
necessary for 4U 1820-30 to be considered a physically relevant
system. According to different investigations done by us, multiple
anisotropic structures meet all physically acceptable conditions
across various metric ansatz \cite{8b}-\cite{8l}. It has been
revealed that the EoS for neutron stars failed to accurately predict
the compactness of some stars like SAX J 1808.4-3658, Her X-1, and
4U 1820-30, etc. Hence, there is a need for another EoS that does
not fail to do so. In this regard, MIT bag model is of great
significance which helps in predicting the compactness of the above
star models correctly \cite{9}-\cite{9f}.

The MIT bag model has been instrumental in probing the intricate
details of quark structures. This model suggests that as the bag
constant, which represents the energy density required to confine
quarks, takes on larger values, the pressure exerted by quarks
decreases correspondingly. Extensive research has been conducted
using this EoS to shed light on the internal composition of strange
objects. In a study of the extremely dense pulsar PSR J 1614-2230,
Demorest et al. \cite{9a} found that the MIT bag model provided a
suitable framework for modeling and understanding the properties of
such ultra-dense stellar objects. Rahaman et al. \cite{10} employed
an interpolation function to estimate the mass of a hypothetical
quark star, and then leveraged this result to elucidate various
physical phenomena associated with such compact objects.
Additionally, Bhar \cite{11} investigated the necessary conditions
for physically realistic stellar models by adopting the MIT EoS in
conjunction with the Krori-Barua ansatz. Deb et al. \cite{12,12a}
explored distinct uncharged and charged fluid models using this bag
model. Similarly, Sharif and his colleagues \cite{13a,13b} adopted
the same approach to analyze stable anisotropic configurations.

A widely accepted theory among scientists, named the general
relativity (GR), has been put forwarded by Albert Einstein in 1915.
The proposal of this fundamental theory is thought to be his one of
the biggest accomplishments. Such a highly recognized theory over
more than one hundred years provides a cornerstone to discuss
stellar and interstellar bodies in relation with their gravitational
interaction at enormous scales. Moreover, this theory has opened
many pathways in detecting mysterious celestial objects.
Nevertheless, some inconsistencies occur in this theory when
researchers explored the origin of cosmic dark components such as
dark energy and dark matter. It is important to know that the former
element produces large enough pressure opposite to the gravity that
leads to the current cosmic rapid expansion era. In order to deal
with such issues, some alternate proposals to GR have been
established in recent years through the modification of the
Einstein-Hilbert action function. The theory, refers to the
first-ever modification of GR is known as the $f(R)$ gravity which
was accomplished through the replacement of $R$ (the Ricci scalar)
with its generic functional. Several $f(R)$ gravity models have been
used to examine different cosmic eras \cite{32}-\cite{32b}. In
addition, using a class of different strategies, a large number of
acceptable astrophysical results have been made in the literature in
the context of $f(R)$ theory \cite{33,33a}.

Many other modifications of GR have been proposed and explored in
different astrophysical as well as cosmological scenarios. Nojiri
and Odintsov \cite{36} contributed in this context and gave a
proposal named as $f(\mathcal{G})$ gravity theory, where
$\mathcal{G}$ symbolizes the Gauss-Bonnet invariant (a sum of
multiple curvature terms). This extension was also produced by
modifying the action, but this time the Ricci scalar is not replaced
by any function rather they added a generic function of
$\mathcal{G}$ in the curvature scalar. Bamba with his collaborators
\cite{14} reconstructed this modified gravity model in order to
discuss several future singularities in relation with
phantom/quintessence cosmic phase. Myrzakulov and his colleagues
\cite{15} solved $f(\mathcal{G})$ gravitational equations without
including the cosmological constant and explored the existence of
anti-gravity force in an inflationary epoch. The expansion of our
universe has been analyzed with the help of some symmetry generators
obtained via Noether symmetry approach in this modified gravity.
Bamba et al. \cite{16} considered multiple Gauss-Bonnet gravity
models and calculated some parametric values for Hubble, snap, jerk,
and deceleration parameters, leading to the presence of normal fluid
source. The spherical stellar structures and their geometry have
also been investigated under the formalism of this gravity
\cite{18a}. Some other interesting works are \cite{18b}-\cite{18e}.

Our aim to do this work is to explore the applicability of modified
Gauss-Bonnet gravity models on the existence of anisotropic stellar
structures admitting spherical symmetry. The following lines
indicate the pattern of this article. Some fundamentals to
understand the formulation of this theory and its field equations
are presented in section \textbf{2}. We also assume the MIT bag
model and Tolman IV ansatz to deal with the gravitational equations.
Section \textbf{3} explores the matching conditions which are
necessary in finding the constants in the considered spacetime under
the observational data of multiple distinct stars. A detailed
discussion through graphical interpretation is done is section
\textbf{4} for two different modified models and several parametric
values. Lastly, section \textbf{5} presents some concluding remarks
about what we have achieved.

\section{Modified Gauss-Bonnet Gravity Theory}

The following modification in the Einstein-Hilbert action in the
presence of fluid's Lagrangian density $\texttt{L}_{m}$ leads to the
modifed Gauss-Bonnet gravity theory as
\begin{equation}\label{1}
\mathrm{I}_{f(\mathcal{G})}=\int
\bigg[\frac{1}{16\pi}\big\{R+f(\mathcal{G})\big\}
+\texttt{L}_{m}\bigg]\sqrt{-g}d^{4}x,
\end{equation}
where $g=|g_{\mu\eta}|$ with $g_{\mu\eta}$ being the metric tensor
and $d^{4}x$ represents an infinitesimal four-volume element in the
spacetime manifold. Also, the Lagrangian term appeared in the above
action can be associated with the energy-momentum tensor
representing fluid source in the self-gravitating interior as
\begin{align}\label{1a}
T_{\mu\eta}=g_{\mu\eta}\texttt{L}_{m}
-\frac{2\partial\texttt{L}_{m}}{\partial g^{\mu\eta}}.
\end{align}
In the current setup, we derive the following gravitational
equations through implementing the variational principle on the
action \eqref{1} given by
\begin{align}\nonumber
&G_{\mu\eta}+(\mathcal{G}f_{\mathcal{G}}-f)g_{\mu\eta}+8\big[R_{\mu\varpi\eta\xi}
+R_{\varpi\eta}g_{\xi\mu}-R_{\varpi\xi}g_{\mu\eta}-R_{\mu\eta}g_{\varpi\xi}\\\label{4a}
&+R_{\mu\xi}g_{\mu\varpi}\big]\nabla^{\varpi}\nabla^{\xi}f_{\mathcal{G}}+4R(g_{\mu\eta}g_{\varpi\xi}
-g_{\mu\varpi}g_{\eta\xi})\nabla^{\varpi}\nabla^{\xi}f_{\mathcal{G}}=8\pi
T_{\mu\eta},
\end{align}
where geometrical aspects of the considered spacetime structure are
exposed by the Einstein tensor $G_{\mu\eta}$. Further,
$\nabla_{\mu}$ is the mathematical expression for the covariant
derivative. When the function $f(\mathcal{G})$ is differentiated
w.r.t. $\mathcal{G}$, we represent it by $f_{\mathcal{G}}$. Once
this functional is removed from the action \eqref{1}, we are left
with the GR framework. One can write Eq.\eqref{4a} in an alternative
way as follows
\begin{equation}\label{4ab}
G_{\mu\eta}=8\pi T^{\textsf{(cor)}}_{\mu\eta}.
\end{equation}
Here, the term on right side represents total fluid distribution and
is classified as follows
\begin{align}\nonumber
T^{\textsf{(cor)}}_{\mu\eta}&=T_{\mu\eta}-(\mathcal{G}f_{\mathcal{G}}-f)g_{\mu\eta}-\frac{1}{\pi}
\big[R_{\mu\varpi\eta\xi}
+R_{\varpi\eta}g_{\xi\mu}-R_{\varpi\xi}g_{\mu\eta}\\\label{4}
&-R_{\mu\eta}g_{\varpi\xi}+R_{\mu\xi}g_{\mu\varpi}\big]\nabla^{\varpi}\nabla^{\xi}f_{\mathcal{G}}
+\frac{R}{2\pi}(g_{\mu\eta}g_{\varpi\xi}-g_{\mu\varpi}g_{\eta\xi})\nabla^{\varpi}\nabla^{\xi}f_{\mathcal{G}},
\end{align}
where $T_{\mu\eta}$ is the usual matter existing in any celestial
body and all other quantities appear due to the modification of
action \eqref{1}.

In the presence of strong gravitational fields, such as around
compact objects like neutron stars, the energy-momentum tensor can
exhibit significant pressure anisotropy. Further, these kinds of
fluid play a crucial role in cosmological models. In the following,
we express such matter distributions by \cite{18f,18g}
\begin{equation}\label{5}
T_{\mu\eta}=(P_{\bot}+\rho)
\mathcal{S}_{\mu}\mathcal{S}_{\eta}-P_{\bot}g_{\mu\eta}
-(P_{\bot}-P_{r})\mathcal{X}_{\mu}\mathcal{X}_{\eta},
\end{equation}
where $\mathcal{S}_{\mu}$ and $\mathcal{X}_{\mu}$ are the
four-vector and four-velocity in the covariant form, satifying
$\mathcal{X}^{\mu}\mathcal{S}_{\mu}=0$ and
$\mathcal{S}^{\mu}\mathcal{S}_{\mu}=1$. Furthermore, the triplet
$(\rho,P_{\bot},P_{r})$ symbolize the energy density, tangential and
radial pressures, respectively. The spherical line element encodes
the symmetries of the spacetime and allows modeling gravitational
fields around compact objects. It is a crucial tool for studying GR
or modified theories in spherical symmetry. The interior region of
such a geometry is represented by
\begin{equation}\label{6}
ds^{2}=e^{\nu}dt^{2}-e^{\lambda}dr^{2}-r^{2}(d\theta^{2}+{\sin^{2}\theta}{d\phi^2}),
\end{equation}
where $\lambda=\lambda(r)$ and $\nu=\nu(r)$, showing that this
metric is of static nature. The quantities appeared in the
energy-momentum tensor \eqref{5} are now become in the light of
metric \eqref{6} as
\begin{equation}\label{6a}
\mathcal{S}^{\mu}=\big(e^{\frac{-\nu(r)}{2}},0,0,0\big),\quad
\mathcal{X}^{\mu}=\big(0,e^{\frac{-\lambda(r)}{2}},0,0\big).
\end{equation}

The methodology utilizing the MIT bag model EoS (describing the
quark's interior) alongside Tolman IV ansatz (which we shall take
into account later) is a widely recognized approach for studying
quark stars. However, the current study offers a new perspective by
applying modified $f(\mathcal{G})$ gravity, which has not been
previously explored in the context of quark stellar structures.
Adopting specific modified gravity models allow researchers to
explore deviations from standard GR and assess their impact on the
behavior of gravity. Moreover, they are used for the interpretation
of observational data more efficiently. In this regard, we employ
two models in the framework of $f(\mathcal{G})$ gravity.
\begin{itemize}
\item \textbf{Model 1:} A quadratic model used to understand the impact of bag
constant in this theory is given as
\begin{equation}\label{2}
f(\mathcal{G})=\beta \mathcal{G}^2,
\end{equation}
where $\beta$ represents an arbitrary constant.
\item \textbf{Model 2:} Another model is
also considered as a realistic one, and is being able to describe
the current cosmic expansion \cite{14}
\begin{equation}\label{3}
f(\mathcal{G})=\delta_{1}\mathcal{G}^{x}(\delta_{2}\mathcal{G}^{y}+1),
\end{equation}
where $x>0$, and $\delta_{1}, \delta_{2}$ and $y$ are free
parameters.
\end{itemize}
This combination enables a more detailed exploration of how
higher-order curvature terms influence the properties of stars,
especially under intense gravitational fields. Hence, this analysis
offers a clearer insight into the behavior of quark stars in the
context of this modified gravity theory. In the context of our
chosen models, the issue of ghost modes emerges due to the
higher-order derivatives introduced by the functional form. These
higher-order terms result in fourth-order equations of motion, which
are typically linked to ghost modes. Although the presence of ghosts
presents a theoretical challenge, these models still play an
important role in advancing our understanding of the behavior of
modified gravitational theories. They underscore the necessity for
more refined models to tackle both cosmological challenges and
ensure theoretical consistency. Nonetheless, they remain significant
in the ongoing exploration of alternative gravitational theories.
Using Eqs.\eqref{4ab} and \eqref{5} along with the metric \eqref{6}
and model 1 yield the field equations as
\begin{align}\label{8}
\rho&=\frac{e^{-2 \lambda}}{8 \pi r^2}\big[\big(e^{\lambda}-1\big)
\big(16 \beta \mathcal{G}''+e^{\lambda }\big)+\lambda '\big\{r
e^{\lambda}-8 \beta \big(e^{\lambda }-3\big)
\mathcal{G}'\big\}+\beta r^2 \mathcal{G}^2 e^{2 \lambda}\big],
\\\label{9}
P_{r}&=\frac{e^{-\lambda}}{8 \pi }\bigg[\frac{\nu '}{{r^2}} \big\{8
\beta \big(3 e^{-\lambda}-1\big)
\mathcal{G}'+r\big\}-e^{\lambda}+1-\beta \mathcal{G}^2
e^{\lambda}\bigg],
\\\nonumber
P_{\bot}&=\frac{e^{-2 \lambda}}{32 \pi r}\big[32 \beta \mathcal{G}'
\nu ''-32 \beta \mathcal{G}'' \nu '+16 \beta \mathcal{G}' \nu '^2-48
\beta \mathcal{G}' \nu '-4 \beta r \mathcal{G}^2 e^{2 \lambda}+2
e^{\lambda} \nu '\\\label{10}&-e^{\lambda} \lambda ' \big(r \nu
'+2\big)-2 r e^{\lambda} \nu ''+r e^{\lambda} \nu '^2\big],
\end{align}
where $'=\frac{\partial}{\partial r}$. It must be stated here that
the field equations corresponding to model 2 are found to be much
lengthy, so we do not include them here. However, the results
corresponding to this model shall be interpreted graphically.
Further, the Gauss-Bonnet term
$$\mathcal{G}=R^{\mu\eta\gamma\delta}R_{\mu\eta\gamma\delta}
+R^2-4R^{\mu\eta}R_{\mu\eta},$$ leads to the following expression
expressed by
\begin{align}\label{10a}
\mathcal{G}&=\frac{2 e^{-2 \lambda}}{r^2}
\big[\big(e^{\lambda}-3\big) {\nu'} {\lambda'}-2
\big(e^{\lambda}-1\big) {\nu''}-\big(e^{\lambda}-1\big)
{\nu'}^2\big],
\end{align}
whose derivatives up to second order are given in Appendix
\textbf{A}. When inserting the overhead values into
Eqs.\eqref{8}-\eqref{10}, the matter variables take the form
presented in Appendix \textbf{A}.

It has been already discussed the importance of some particular
constraints, say an EoS that can help in understanding the features
of consider structures in a more better way. The main purpose to
take EoSs into account is that each of them associate different
physical parameters in a certain manner. One of the fascinating
events occurring in our universe is the existence of neutron star
like structure which are the results of collapse of highly massive
stars. Such a structure has much powerful field of attraction around
it with being of small size. Among these objects, the one comprising
of low density converts into the quark star. A non-linear system
\eqref{11a}-\eqref{11c} with five unknowns require an EoS to discuss
the quark interior. A suitable candidate in this regard is the MIT
bag model, providing a simple yet effective way to model the
confinement of quarks inside hadrons. By confining the quarks within
a bag with appropriate boundary conditions, the model captures the
essential features of color confinement.

Quark pressure and quark density are important properties of the
quark-gluon plasma. We define the former element in relation with
the bag constant $\mathfrak{B}$ as
\begin{equation}\label{11}
P_{r}=\sum_{\hbar}P^{\hbar}-\mathfrak{B}, \quad
\hbar=\texttt{d},\texttt{u},\texttt{s},
\end{equation}
where we observe that such a matter usually classifies into three
flavors, namely down ($\texttt{d}$), up ($\texttt{u}$) and strange
($\texttt{s}$). In addition, the quark pressures $P^{\texttt{d}}$,
$P^{\texttt{u}}$ and $P^{\texttt{s}}$ correspond to these flavors,
respectively. We now define the quark density as
\begin{equation}\label{12}
\rho=\sum_{\hbar}\rho^{\hbar}+\mathfrak{B}, \quad
\hbar=\texttt{d},\texttt{u},\texttt{s}.
\end{equation}
It is important to mention here that the relation between quark
pressure and density can be seen from $\rho^{\hbar}=3P^{\hbar}$.
Using this along with Eqs.\eqref{11} and \eqref{12} at the same time
leads to the construction of the desired MIT bag model having the
following form
\begin{equation}\label{13}
P_{r}=\frac{1}{3}(\rho-4\mathfrak{B}).
\end{equation}
The literature is full of exploration of certain salient properties
linked with strange quark stars by choosing multiple values of the
bag constant \cite{25}-\cite{25c}. Hence, this constant comprises of
a broader range of values.

Quantum Chromodynamics (QCD) is the fundamental theory that governs
the strong nuclear force, one of the four basic forces of nature. A
key aspect of QCD is confinement, which means that quarks cannot be
found in isolation but are always confined within particles like
protons and neutrons by the strong force. However, under extreme
conditions of temperature and density, such as those in the early
universe or within neutron stars, quarks and gluons can break free
from confinement and form a state known as quark-gluon plasma.
Investigating this plasma state provides crucial insights into how
QCD behaves under such extreme conditions. In astrophysical
contexts, QCD is vital for understanding the properties of
ultra-dense matter inside objects like neutron stars or theoretical
quark stars, where quarks may exist in a deconfined state. The MIT
bag model \eqref{13} is a theoretical framework that models quark
matter, incorporating concepts from QCD to describe its behavior.

Plugging the above-mentioned constraint into the field equations
\eqref{11a}-\eqref{11c}, they contain the significant role of the
bag constant. After taking EoS \eqref{13} into account, we have four
equations with six unknowns
($\lambda,\nu,\rho,P_{\bot},P_{r},\mathfrak{B}$). At this stage, the
unique solution to this system is only possible when we adopt a
metric ansatz. Therefore, in order to make it sure, we adopt
non-singular Tolman IV metric potentials \cite{40}, described as
\begin{equation}\label{14}
e^{\nu}=\left(1+\frac{r^2}{\mathbb{A}^2}\right)\mathbb{B}^{2},
\quad\quad
e^{\lambda}=\frac{1+\frac{2r^2}{\mathbb{A}^2}}{\left(1-\frac{r^2}{\mathbb{C}^2}\right)
\left(1+\frac{r^2}{\mathbb{A}^2}\right)},
\end{equation}
involving a constant triplet
($\mathbb{A}^2,\mathbb{B}^2,\mathbb{C}^2$). The Appendix \textbf{B}
contains the values of fluid parameters after substituting the
ansatz \eqref{14} into them.

\section{Boundary Conditions}

Boundary conditions are important in providing better understanding
of the structural development of any celestial body. These are the
constraints that help in matching the two regions, i.e., interior
and exterior of a compact star at some boundary, named as
hypersurface. Since the interior spacetime is a static sphere, it is
better to adopt the Schwarzschild metric as an exterior geometry.
This metric is defined as follows
\begin{equation}\label{15}
ds^2=\bigg(1-\frac{2\mathcal{M}}{r}\bigg)dt^2
-\bigg(1-\frac{2\mathcal{M}}{r}\bigg)^{-1}dr^2
-r^2d\theta^{2}-r^2{\sin^{2}\theta}{d\phi^2},
\end{equation}
where the exterior mass is symbolized by $\mathcal{M}$. In
accordance with the first fundamental form of the boundary
conditions, both $g_{rr}$ and $g_{tt}$ as well as the radial
derivative of the later component corresponding to exterior and
interior geometries are continuous across the spherical interface.
Following this, we get
\begin{align}\label{16}
g_{tt}&=\bigg(1+\frac{\mathcal{R}^2}{\mathbb{A}^2}\bigg)\mathbb{B}^{2}
=1-\frac{2\mathcal{M}}{\mathcal{R}},\\\label{17}
g_{rr}&=\frac{(1-\frac{\mathcal{R}^2}{\mathbb{C}^2})(1+\frac{\mathcal{R}^2}{\mathbb{A}^2})}{1+\frac{2\mathcal{R}^2}{\mathbb{A}^2}}
=1-\frac{2\mathcal{M}}{\mathcal{R}},\\\label{18} \frac{\partial
g_{tt}}{\partial
r}&=\frac{2\mathcal{R}}{\mathbb{A}^2+\mathcal{R}^2}=\frac{2\mathcal{M}}{\mathcal{R}(\mathcal{R}-2\mathcal{M})}.
\end{align}
The above three equations are simultaneously solved, providing the
Tolman IV triplet $(\mathbb{A}^2,\mathbb{B}^2,\mathbb{C}^2)$. We
express them in the following
\begin{equation}\label{19}
\mathbb{A}^2=\frac{\mathcal{R}^2(\mathcal{R}-3\mathcal{M})}{\mathcal{M}},\quad
\mathbb{B}^2=\frac{\mathcal{R}-3\mathcal{M}}{\mathcal{R}},\quad
\mathbb{C}^2=\frac{\mathcal{R}^3}{\mathcal{M}}.
\end{equation}

It is well recognized outcome in the study of compact stars that the
radial pressure must be a decreasing function of the radial
coordinate and disappears for maximum possible value of $r$, i.e.,
at the interface. We use this condition for the radial pressure
(presented in Appendix \textbf{B}) corresponding to model 1 as
\begin{align}\nonumber
P_{r}\mid_{r={\mathcal{R}}}&=\frac{1}{12 \pi \mathcal{R}^{12}
(\mathcal{M}-\mathcal{R})^3}\{12 \mathcal{M}^2 \big(4 \pi
\mathfrak{B} \mathcal{R}^{13}+\mathcal{R}^{11}\big)-\mathcal{M}^3
\mathcal{R}^{10} \big(16 \pi \mathfrak{B}
\mathcal{R}^2\\\nonumber&+15\big)-3 \mathcal{M} \big(16 \pi
\mathfrak{B} \mathcal{R}^{14}+\mathcal{R}^{12}\big)+16 \pi
\mathfrak{B} \mathcal{R}^{15}+1452672 \beta
\mathcal{M}^7-\beta\mathcal{R}\\\label{22}&\times 1388160
\mathcal{M}^6+395136 \beta \mathcal{M}^5 \mathcal{R}^2+6
\mathcal{M}^4 \big(\mathcal{R}^9-5184 \beta
\mathcal{R}^3\big)\}^{-1}.
\end{align}
The above equation leads to the following value of bag constant for
model 1
\begin{align}\nonumber
\mathfrak{B}&=\frac{1}{48 \pi\mathcal{M} \mathcal{R}^{12}
(\mathcal{M}-\mathcal{R})^3}\big\{484224 \beta \mathcal{M}^6-462720
\beta  \mathcal{M}^5 \mathcal{R}-5 \mathcal{M}^2
\mathcal{R}^{10}\\\label{23}&+131712 \beta \mathcal{M}^4
\mathcal{R}^2+2 \mathcal{M}^3 \big(\mathcal{R}^9-5184 \beta
\mathcal{R}^3\big)+4 \mathcal{M}
\mathcal{R}^{11}-\mathcal{R}^{12}\big\}^{-1},
\end{align}
whereas for model 2, the bag constant is not specified here due to
lengthy expression. Five different strange quark candidates are
chosen along with their observational data (masses and radii) and we
present them in Table \textbf{1} along with their compactness. It is
important to note that the limit for this factor has been suggested
by a researcher, referred to the Buchdhal limit \cite{22c} defined
as $\frac{2\mathcal{M}}{\mathcal{R}}<\frac{8}{9}$. From Table
\textbf{1}, we see that the considered stars are in agreement with
this limit. We use this experimental data to compute the unknown
constants, i.e., $\mathbb{A}^2,\mathbb{B}^2$ and $\mathbb{C}^2$ and
present them in Table \textbf{2}. The bag constant is also
numerically determined in Tables \textbf{3} and \textbf{4}
corresponding to models 1 and 2, respectively.
\begin{table}[H]
\scriptsize \centering \caption{Observational data for different
strange quark stars.} \label{Table1} \vspace{+0.1in}
\setlength{\tabcolsep}{0.95em}
\begin{tabular}{cccccc}
\hline\hline \\Strange Stars & Her X-I & LMC X-4 & PSR J 1614-2230 &
Vela X-I &4U 1820-30
\\\hline \\
Mass $(\mathcal{M}_{\odot})$ & 0.85 & 1.04 & 1.97 & 1.77 & 1.58
\\\hline \\
$\mathcal{R}$ (km) & 8.1 & 8.3 & 9.69 & 9.56 & 9.1
\\\hline \\
$\mathcal{M}/\mathcal{R}$ & 0.1546 & 0.1846 & 0.2996 & 0.2728 & 0.2559 \\
\hline\hline
\end{tabular}
\end{table}
\begin{table}[H]
\scriptsize \centering \caption{Numerically calculated Tolman IV
triplet ($\mathbb{A}^2$, $\mathbb{B}^2$, $\mathbb{C}^2$).}
\label{Table1} \vspace{+0.1in} \setlength{\tabcolsep}{0.95em}
\begin{tabular}{cccccc}
\hline\hline \\Strange Stars & Her X-I & LMC X-4 & PSR J 1614-2230 &
Vela X-I & 4U 1820-30
\\\hline \\
$\mathbb{A}^2$ & 227.339 & 166.325 & 32.0351 & 62.6318 & 75.141
\\\hline \\
$\mathbb{B}^2$ & 0.535963 & 0.445918 & 0.101924 & 0.184694
& 0.232224\\\hline \\
$\mathbb{C}^2$ & 424.169 & 372.995 & 314.305 & 339.112 & 323.571 \\
\hline\hline
\end{tabular}
\end{table}
\begin{table}[H]
\scriptsize \centering \caption{Numerically calculated bag constant
for model 1.} \label{Table3} \vspace{+0.1in}
\setlength{\tabcolsep}{0.9em}
\begin{tabular}{cccccc}
\hline\hline \\Strange Stars & Her X-I & LMC X-4 & PSR J1614-2230&
Vela X-I & 4U 1820-30
\\\hline \\
$\mathfrak{B}$ ($\beta$=0.5) & 0.000114958 & 0.000123759
& 0.000108751 & 0.000110312 & 0.000121002\\
\hline\\
$\mathfrak{B}$ ($\beta$=5.5) & 0.000114947 & 0.000123724
& 0.000108699 & 0.000110255 & 0.000120925\\
\hline\hline
\end{tabular}
\end{table}
\begin{table}[H]
\scriptsize \centering \caption{Numerically calculated bag constant
for model 2.} \label{Table4} \vspace{+0.1in}
\setlength{\tabcolsep}{0.75em}
\begin{tabular}{cccccc}
\hline\hline \\Strange Stars & Her X-I & LMC X-4 & PSR J1614-2230&
Vela X-I & 4U 1820-30
\\\hline \\
$\mathfrak{B}$ ($\delta_{1}$=0.2, $\delta_{2}$=0.5) & 0.000114958 &
0.000123759 & 0.000110935 & 0.000110312 & 0.000121002\\\hline \\
$\mathfrak{B}$ ($\delta_{1}$=0.2, $\delta_{2}$=8.5) & 0.000114956 &
0.000123751 & 0.000108739 & 0.000110298 & 0.000120983\\\hline \\
$\mathfrak{B}$ ($\delta_{1}$=5, $\delta_{2}$=0.5) & 0.000114954 &
0.000123745 & 0.00010873 & 0.000110289 & 0.000120971\\\hline \\
$\mathfrak{B}$ ($\delta_{1}$=5, $\delta_{2}$=8.5) & 0.000114865 &
0.000123466 & 0.000108315 & 0.000109835 & 0.000121363\\
\hline\hline
\end{tabular}
\end{table}

The range of the bag constant is found to be within the interval
$(58.9,91.5)$ $MeV/fm^3$ for the strange compact bodies having not
sufficient mass \cite{26}. However, when the quark bodies having
mass approximately equal to 154 $MeV$ are concerned, the above range
becomes 56 to 78 $MeV/fm^3$ \cite{27}. The literature stresses
multiple works on the use of bigger values of this constant from
which several remarkable results have been achieved. In the light of
this, it was shown by Xu \cite{28} that a particular celestial
structure, namely LMXB EXO 0748-676 behaves like a quark like star
when choosing $\mathfrak{B}=60$ and $100~MeV/fm^3$. Some other
experiments, in particular, CERN-SPS and RHIC, suggested that the
quantity $\mathfrak{B}$ can take a more wider range for the
density-dependent structures. The values of bag constant (in
$MeV/fm^{3}$) corresponding to the strange stars Her X-I, Vela X-I,
PSR J1614-2230, LMC X-4 and 4U 1820-30, respectively, for model 1
with mentioned parametric values come out to be
\begin{itemize}
\item $\beta=0.5$: the values are
86.87, 83.36, 82.18, 93.52 and 91.43.
\item $\beta=5.5$: the values are 86.86, 83.31,
82.14, 93.49 and 91.38.
\end{itemize}
For model 2, they are
\begin{itemize}
\item $\delta_{1}=0.2,~\delta_{2}=0.5$: the
values are 86.86, 83.35, 83.82, 93.51 and 91.42.
\item $\delta_{1}=0.2,~\delta_{2}=8.5$: the
values are 86.86, 83.34, 82.16, 93.51 and 91.41.
\item $\delta_{1}=5,~\delta_{2}=0.5$: the values
are 86.86, 83.33, 82.16, 93.50 and 91.40.
\item $\delta_{1}=5,~\delta_{2}=8.5$: the values are 86.79,
82.99, 81.84, 93.29 and 90.95.
\end{itemize}

\section{Graphical Exploration of Strange Quark Stars}

In this section, we illustrate the graphical plots to elucidate the
significant impact of $f(\mathcal{G})$ gravity on certain
astronomical structures. To delve into their fundamental attributes,
we focus on the masses and radii of constructed quark entities. The
$f(\mathcal{G})$ models facilitate this exploration by choosing
different values of the constants involved in these models. The
stars are graphically interpreted on the basis of the following
selected values of the constants in the models
\begin{itemize}
\item For model 1, two values are chosen as $\beta=0.5$ and $5.5$.
\item For model 2, two values of both parameters are adopted as
$\delta_{1}=0.2,~5$ and $\delta_{2}=0.5,~8.5$ along with $x=1=y$.
\end{itemize}
Our analysis commences with an examination of the viability of the
temporal and radial metric functions. Subsequently, we plot the
matter variables, anisotropy, and energy conditions. Following this,
we employ two distinct criteria to assess the stability of the
assumed stars. It is crucial to note that the plots of metric
functions should exhibit an increasing trend with $r$ to ensure the
compatibility of the solution. The metric coefficients, represented
by Eq.\eqref{14}, involve three unknowns
($\mathbb{A}^2,\mathbb{B}^2,\mathbb{C}^2$), whose values are
provided in Table \textbf{2}. Additionally, the compatibility of the
solution can be verified from Figure \textbf{1}.
\begin{figure}\center
\epsfig{file=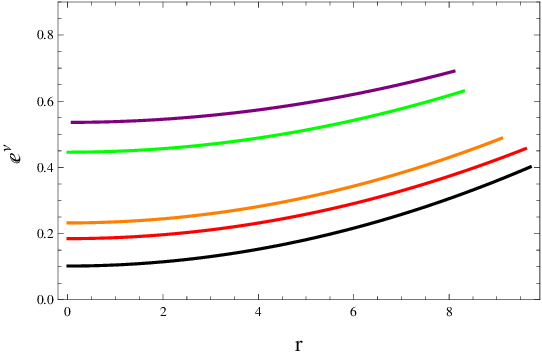,width=0.45\linewidth}\epsfig{file=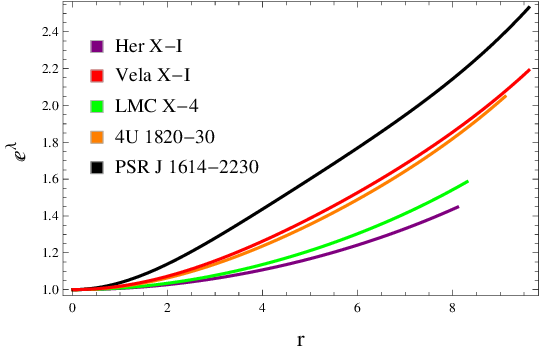,width=0.45\linewidth}
\caption{Tolman IV metric potentials.}
\end{figure}

\subsection{Profile of Physical Determinants}
\begin{figure}[h!]\center
\epsfig{file=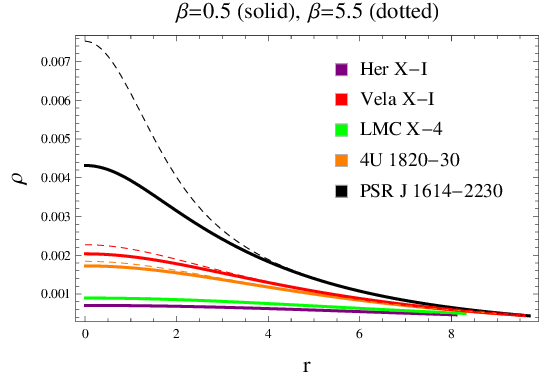,width=0.35\linewidth}\epsfig{file=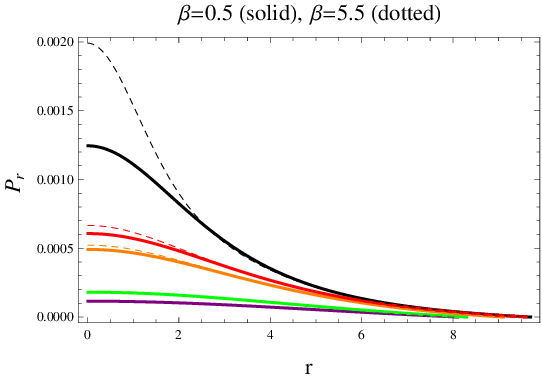,width=0.35\linewidth}\epsfig{file=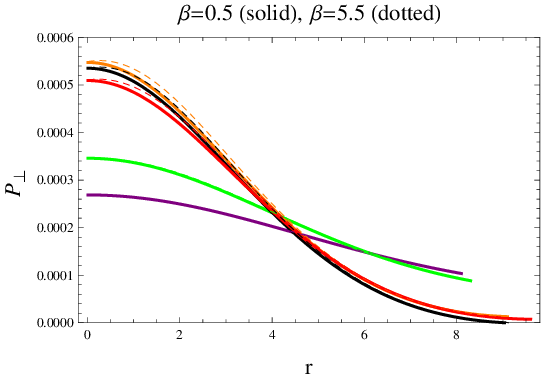,width=0.35\linewidth}
\epsfig{file=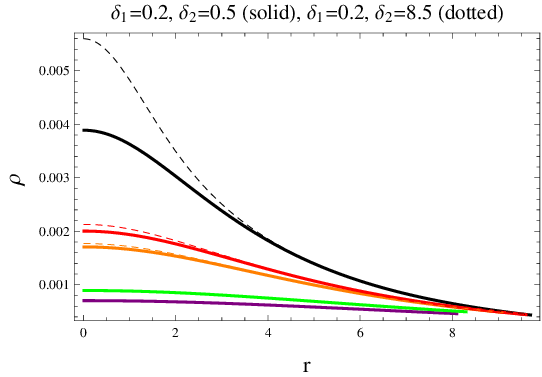,width=0.35\linewidth}\epsfig{file=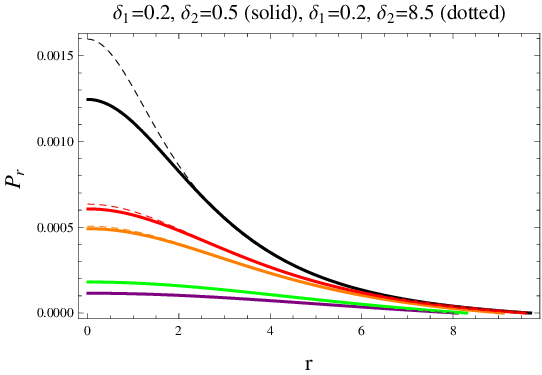,width=0.35\linewidth}\epsfig{file=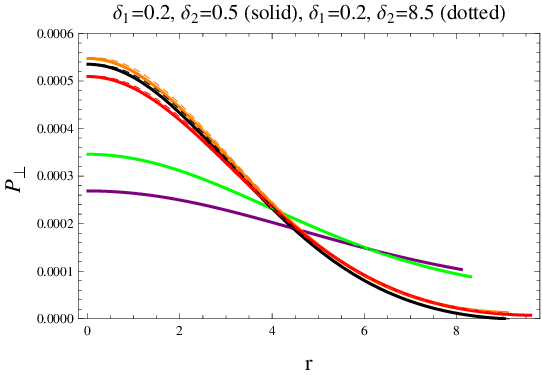,width=0.35\linewidth}
\epsfig{file=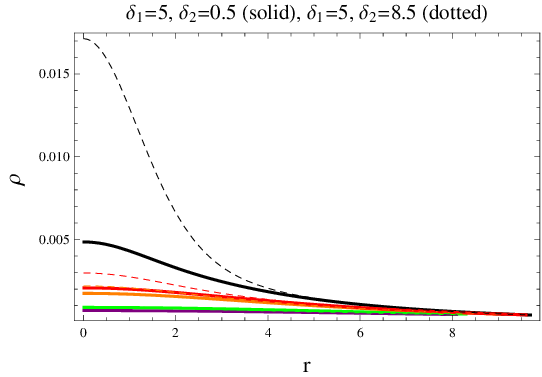,width=0.35\linewidth}\epsfig{file=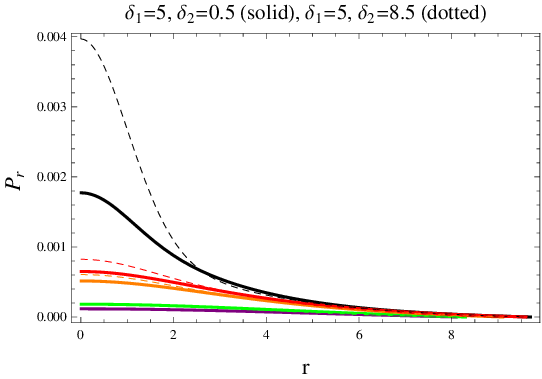,width=0.35\linewidth}\epsfig{file=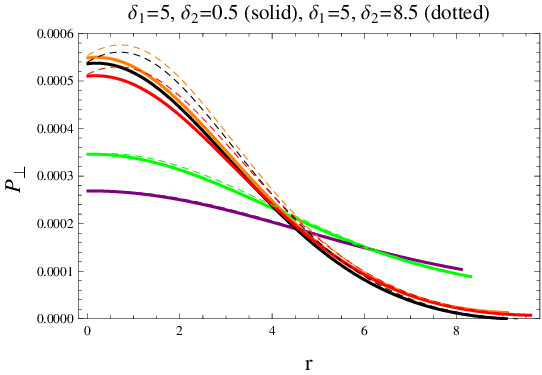,width=0.35\linewidth}
\caption{Fluid parameters for model 1 (first row) and model 2
(second and third rows).}
\end{figure}

The significance of energy density and pressure components within
anisotropic stars cannot be overlooked as they play pivotal roles in
the structural evolution. The feasibility of these compact bodies
hinges greatly upon the behavior of these physical parameters.
Specifically, if these factors exhibit maximum values at the center,
maintain a positive trend throughout, and smoothly decrease towards
the boundary, then such behavior ensure the regularity of the
models. Notably, the density of the system peaks at the central
point for all stars and decreases as the radial coordinate, denoted
by $r$, increases (see Figure \textbf{2}). This trend is also
mirrored by the tangential and radial pressures, as illustrated in
Figure \textbf{2}, with $P_{r}$ reaching to zero at the boundary.
Additionally, specific constraints, referred to the maximality
conditions must be satisfied to ensure the acceptability of the
solution, given as
\begin{align*}
\frac{d\rho}{dr}|_{r=0}=0=\frac{dP_{r}}{dr}|_{r=0}, \quad
\frac{d^2\rho}{dr^2}|_{r=0}<0,\quad \frac{d^2P_{r}}{dr^2}|_{r=0}<0.
\end{align*}
All the requirements for the regularity of solution are satisfied by
every considered candidate for all the mentioned values of both
models' parameters. However, their plots are not added.

\subsection{Anisotropic Factor}

Anisotropy is created when the system has different pressures in
different directions (radial and transversal), expressed by
$\Delta=P_{\bot}-P_{r}$. The pressure ingredients presented in
Appendix \textbf{B} are utilized to compute the anisotropic factor
for model 1 in terms of bag constant and Tolman IV solution, and is
given in the same Appendix. For model 2, the corresponding
expression is not included here. The structural changes in the
strange candidates produced as a result of anisotropy are observed
by utilizing the observational data of stars, as provided in Table
\textbf{1}. The anisotropy of the system in which the pressure in
the radial direction is smaller than the tangential direction comes
out to be positive. Under such circumstances, repulsive pressure is
released to make the structure stable. On the other hand, attractive
force is observed for the system having larger radial pressure than
tangential pressure. In our current setup, the anisotropy in Figure
\textbf{3} shows the positive behavior for all the considered quark
stars.
\begin{figure}\center
\epsfig{file=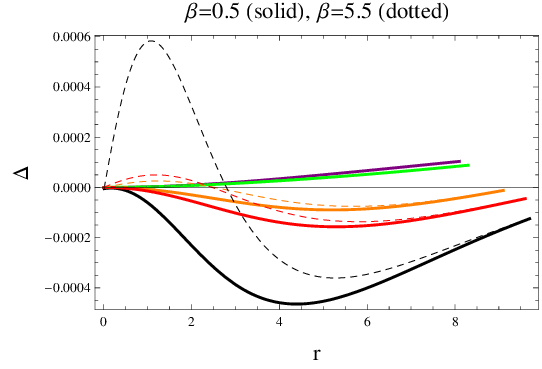,width=0.35\linewidth}\epsfig{file=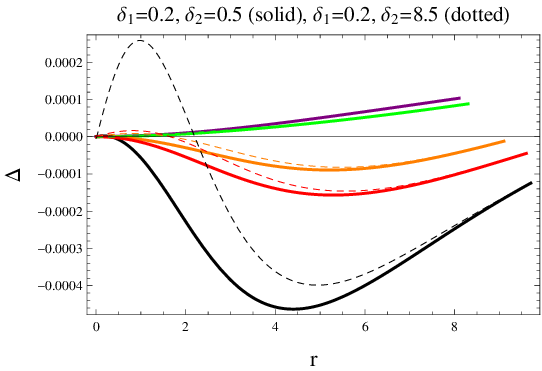,width=0.35\linewidth}\epsfig{file=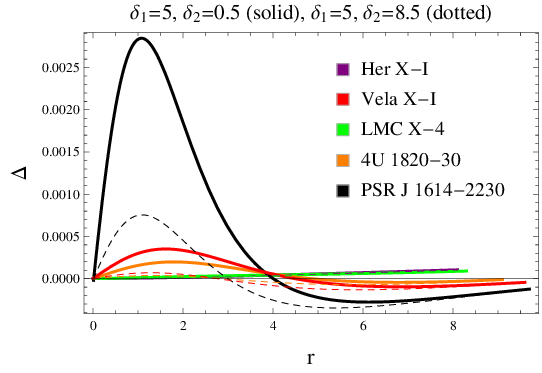,width=0.35\linewidth}
\caption{Anisotropy for model 1 (left) and model 2 (mid and right).}
\end{figure}

\subsection{Energy Conditions}

To ensure the existence of usual matter configuration within the
interiors of the celestial bodies, several mathematical constraints
are considered. These constraints, known as energy conditions, serve
as effective means to distinguish between usual and exotic fluid
sources. Fulfillment of these conditions validates the viability of
the derived solution and confirms the presence of ordinary fluid
within compact structures. In cases of anisotropic composition,
these conditions are categorized into strong, dominant, null, and
weak energy conditions
\begin{align}\nonumber
\text{Strong energy conditions:}\quad &\rho+P_{r}+2P_{\bot}\geq0,
\\\nonumber\text{Dominant energy conditions:}\quad &\rho
-P_{\bot}\geq0,\quad \rho-P_{r}\geq0,
\\\nonumber\text{Null energy conditions:}\quad&\rho
+P_{r}\geq0,\quad\rho +P_{\bot}\geq0,
\\\label{59} \text{Weak energy conditions:}\quad &
\rho+P_{r}\geq0,\quad\rho +P_{\bot}\geq0.
\end{align}
Figures \textbf{4}-\textbf{6} illustrate the energy plots depicting
all compact entities under investigation, each corresponding to
chosen values of prescribed models. Analysis of these plots reveals
that all imposed constraints are effectively met, affirming the
viability of the proposed solutions. Consequently, it can be
inferred that all quark entities contain ordinary sources within
their structures, thus indicating a coherent and viable model within
this modified theoretical framework.
\begin{figure}\center
\epsfig{file=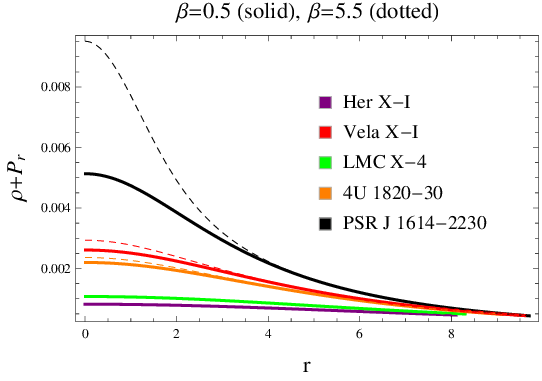,width=0.4\linewidth}\epsfig{file=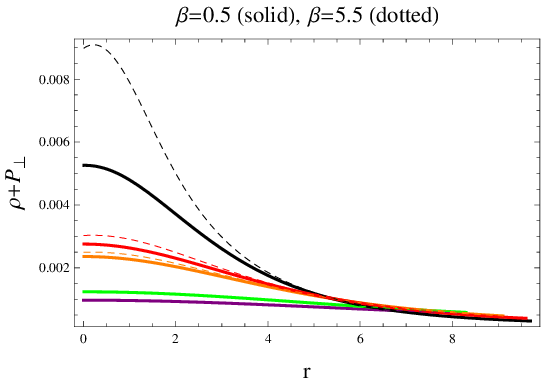,width=0.4\linewidth}
\epsfig{file=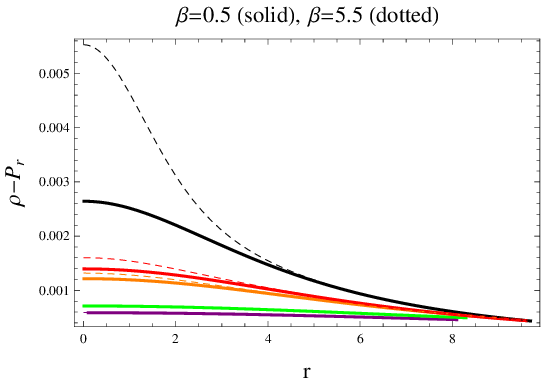,width=0.4\linewidth}\epsfig{file=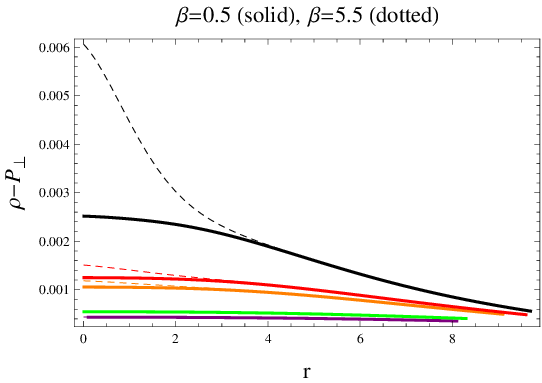,width=0.4\linewidth}
\epsfig{file=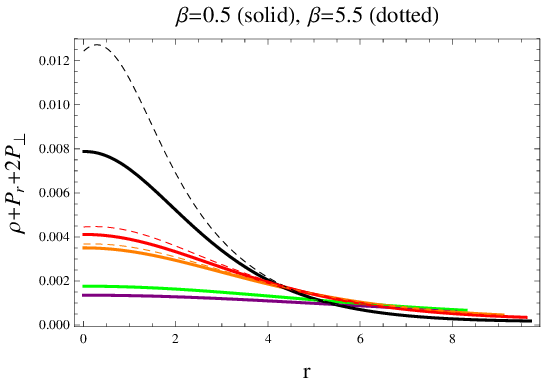,width=0.4\linewidth} \caption{Energy
conditions for model 1.}
\end{figure}
\begin{figure}\center
\epsfig{file=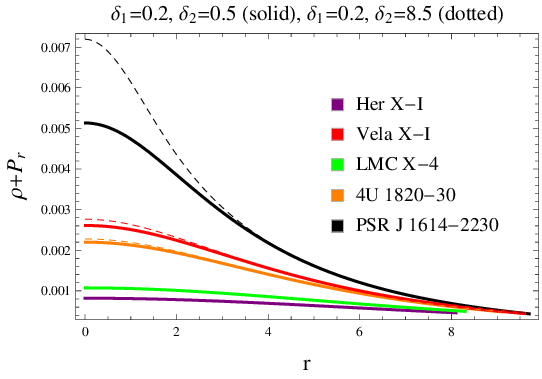,width=0.4\linewidth}\epsfig{file=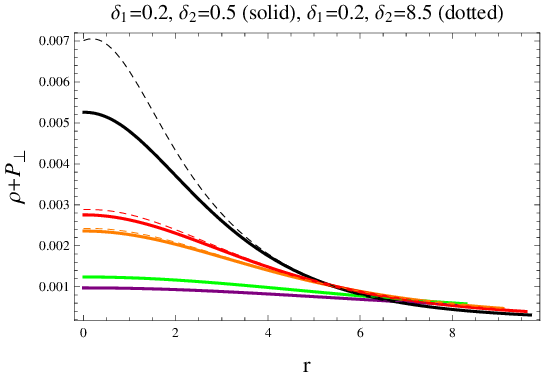,width=0.4\linewidth}
\epsfig{file=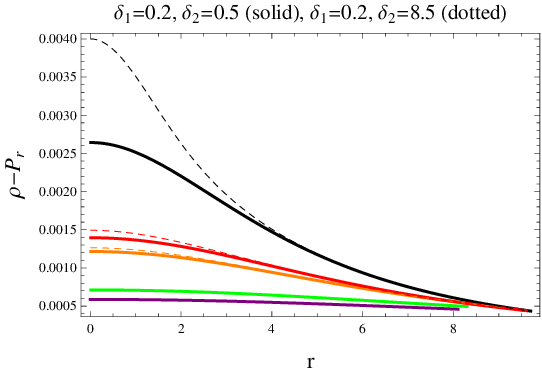,width=0.4\linewidth}\epsfig{file=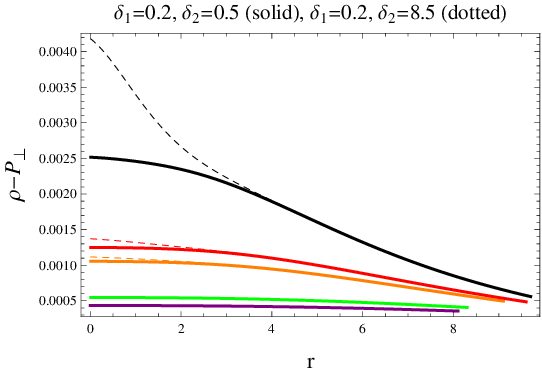,width=0.4\linewidth}
\epsfig{file=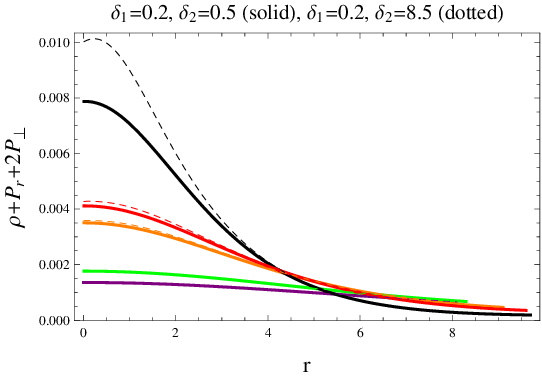,width=0.4\linewidth} \caption{Energy
conditions for model 2.}
\end{figure}
\begin{figure}\center
\epsfig{file=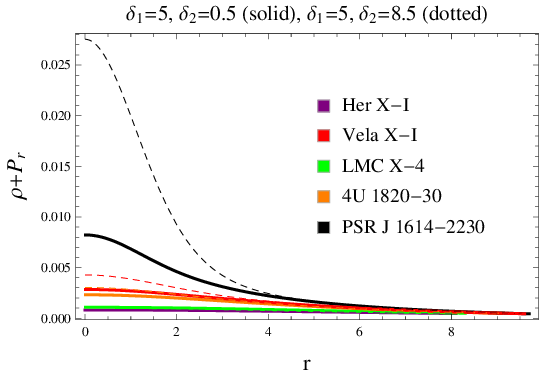,width=0.4\linewidth}\epsfig{file=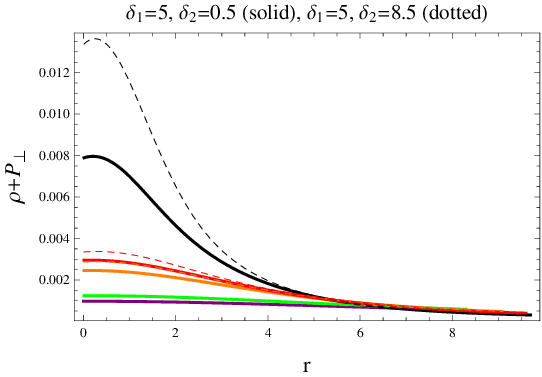,width=0.4\linewidth}
\epsfig{file=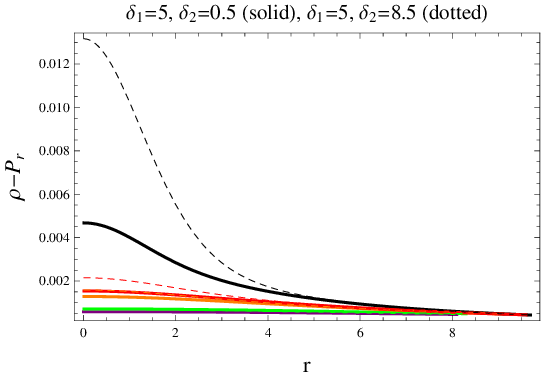,width=0.4\linewidth}\epsfig{file=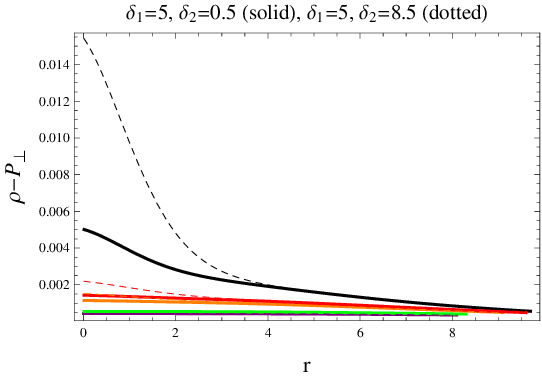,width=0.4\linewidth}
\epsfig{file=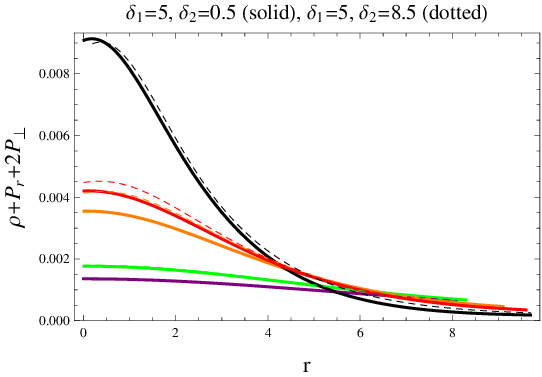,width=0.4\linewidth} \caption{Energy
conditions for model 2.}
\end{figure}

\subsection{Stability Tests}

We shall investigate the stability (which is a significant topic
within the cosmic universe) of spherically symmetric anisotropic
stars in this subsection. This shall be performed through two
distinct methods. At first, the causality condition is utilized as a
mean to understand the stability of stellar bodies by considering
its various components such as
\begin{equation}\label{59a}
\mathcal{V}^{2}_{r}=\frac{dP_{r}} {d\rho},\quad\quad
\mathcal{V}^{2}_{\bot}=\frac{dP_{\bot}} {d\rho},
\end{equation}
where $\mathcal{V}^{2}_{r}$ and $\mathcal{V}^{2}_{\bot}$ signify the
radial and transversal sound speed components, respectively. This
criterion is satisfied when both inequalities
$0\leq\mathcal{V}^{2}_{\bot}\leq1$ and
$0\leq\mathcal{V}^{2}_{r}\leq1$ hold \cite{24ab,24ac}, and
alternatively, one can say that sound speed in every direction must
be less than the speed of light. The other technique used to
determine the stable quarks is adiabatic index, whose expression is
delineated as
\begin{equation}\label{59b}
\Gamma=\frac{\rho+P_{r}}{P_{r}}\bigg(\frac{d P_{r}}{d\rho}\bigg).
\end{equation}
The anisotropic stars depict stable interior if the value of
$\Gamma$ is greater than $\frac{4}{3}$ \cite{25a}. Figure \textbf{7}
represents the plots of the two considered tests for both the
respective models \eqref{2} and \eqref{3}. From this Figure, it can
be easily observed that the model 1 portrays the stable quarks for
both the proposed values of $\beta$. The middle row indicates that
all the five stars corresponding to $\delta_{1}=0.2$ and
$\delta_{2}=0.5$ are stable. While on the contrary, the stars 4U
1820-30, Vela X-1 and PSR J 1614-2230 concerning $\delta_{1}=0.2$
and $\delta_{2}=8.5$ could not fulfill the stability tests and
hence, are unstable. The last row also illustrates the stability
approach for a couple of considered parametric values and one can
notice that all the stars are underlying in the stable regions.
\begin{figure}\center
\epsfig{file=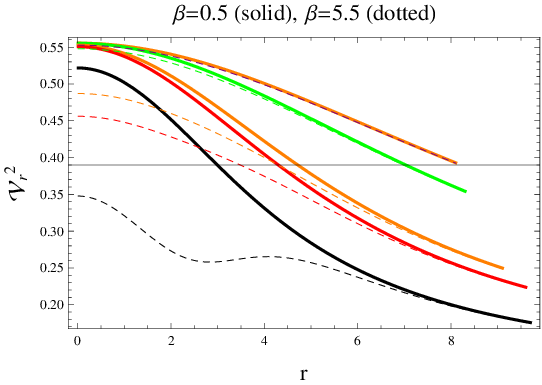,width=0.35\linewidth}\epsfig{file=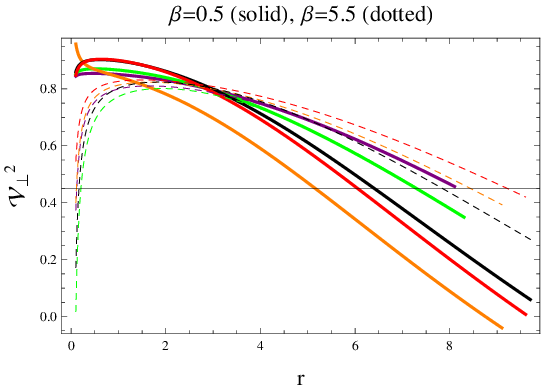,width=0.35\linewidth}\epsfig{file=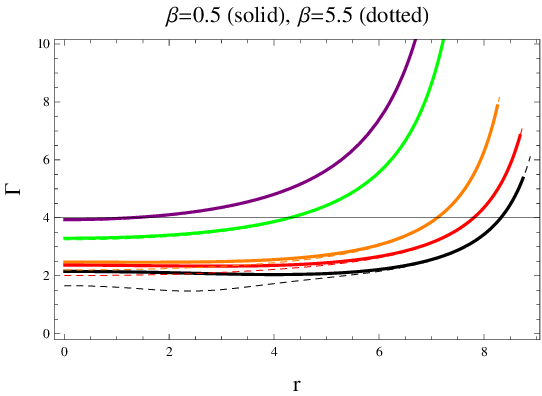,width=0.35\linewidth}
\epsfig{file=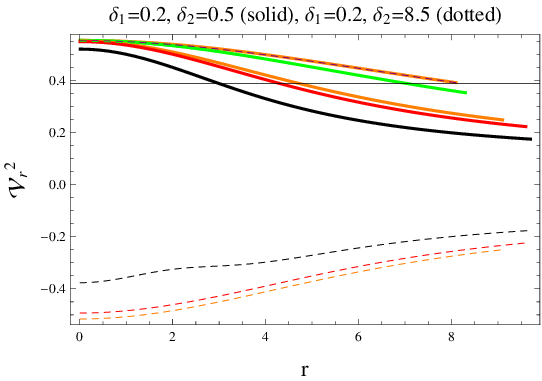,width=0.35\linewidth}\epsfig{file=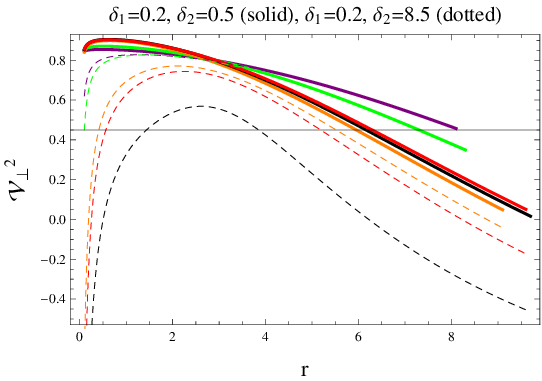,width=0.35\linewidth}\epsfig{file=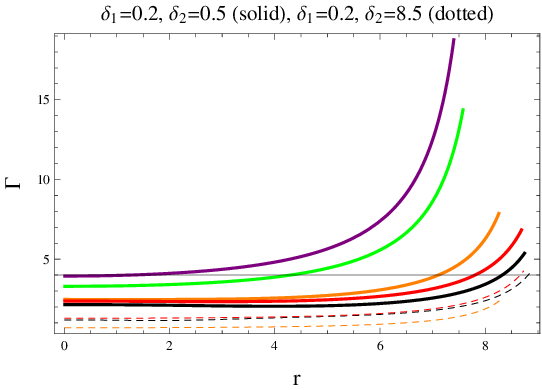,width=0.35\linewidth}
\epsfig{file=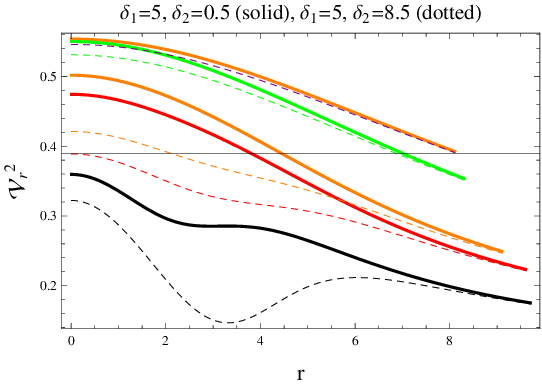,width=0.35\linewidth}\epsfig{file=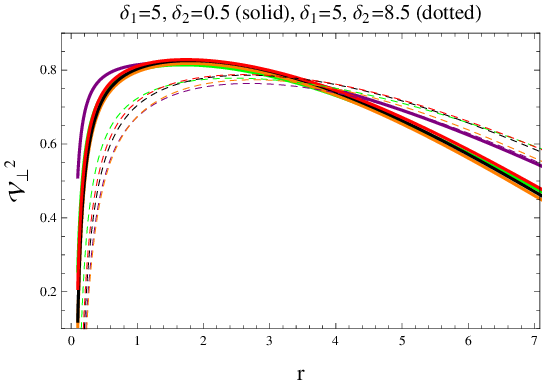,width=0.35\linewidth}\epsfig{file=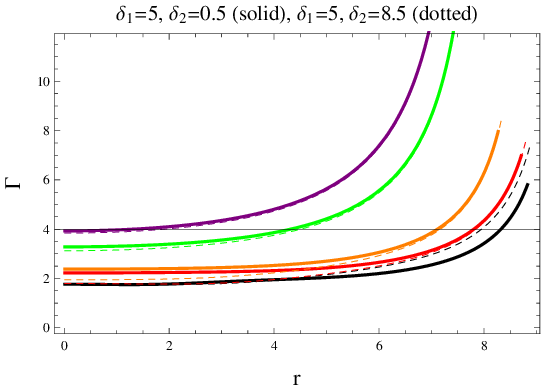,width=0.35\linewidth}
\caption{Stability for model 1 (first row) and model 2 (second and
third rows).}
\end{figure}

\subsection{Some other Essential Factors}

A fundamental quantity linked with the heavily celestial structures
is the mass function. For spherically symmetric configurations, this
function can be calculated through the following integral equation
\begin{equation}\label{58}
\mathit{m}(r)=\int^{\mathcal{R}}_{0}4\pi\rho \bar{r}^2d\bar{r}.
\end{equation}
We utilize a particular initial condition on this function, i.e.,
$\mathit{m}(0)=0$ to solve Eq.\eqref{58}. Keep in mind that we use
the above definition instead of the one in terms of radial metric
potential so that the effect of modification of functional form can
be explored on strange stellar structures. It is important to gather
some information about a factor related to the mass, referred to the
compactness. It is widely used in literature while discussing
compact celestial objects and expressed as a ratio between mass and
radius, i.e., $\psi(r)=\frac{\mathit{m}(r)}{r}$. When matching two
(interior and exterior) regions of a spherical geometry, Buchdahl
\cite{22c} came to know the expected possible value of this factor.
He deduced it not to be more than $\frac{4}{9}$ everywhere to get a
physically existing star model.

Almost all the astrophysical structures produce electromagnetic
beams or radiations when influenced from the strong force of
attraction produced by other nearby objects. These radiations are
increased in terms of their wavelength and this can be found through
the following expression
\begin{equation*}
\sigma=\frac{1}{\sqrt{1-2\psi}}-1.
\end{equation*}
Since this is also dependent on the mass, the graphical behavior of
this factor must be an increasing function of $r$. As its acceptable
value/range is concerned, researchers found this to be less than $2$
only when isotropic fluid configurations are discussed. On the other
hand, when anisotropy in the interior is considered, this value
reaches to $5.211$. The profiles of all the above physical factors
depicted in Figure \textbf{8} show that they are consistent with
their limits for every parametric choice.
\begin{figure}[h!]\center
\epsfig{file=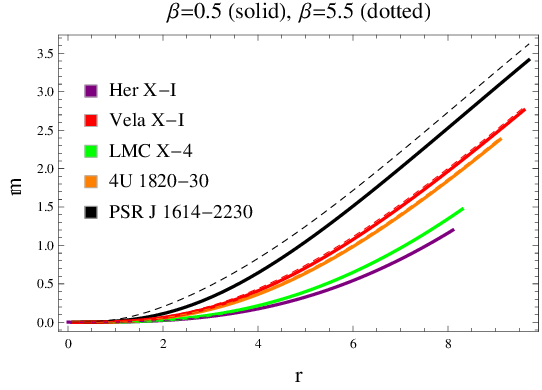,width=0.35\linewidth}\epsfig{file=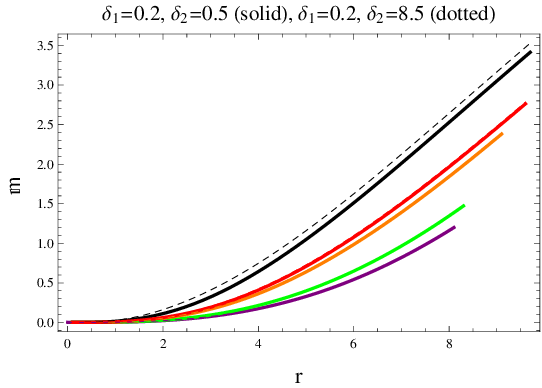,width=0.35\linewidth}\epsfig{file=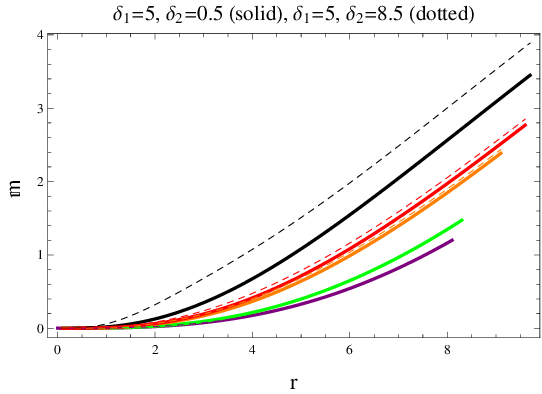,width=0.35\linewidth}
\epsfig{file=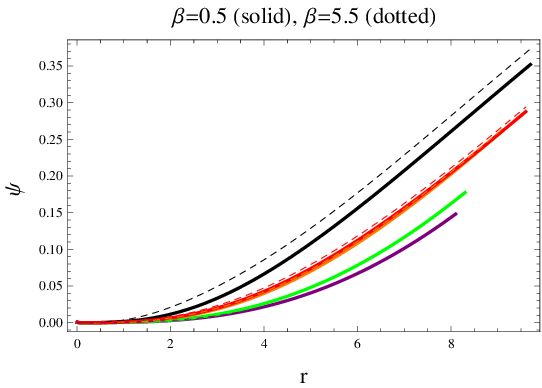,width=0.35\linewidth}\epsfig{file=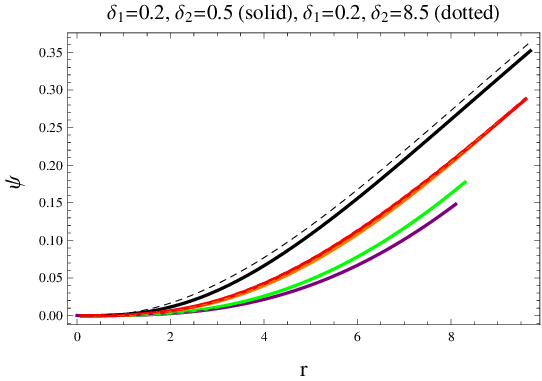,width=0.35\linewidth}\epsfig{file=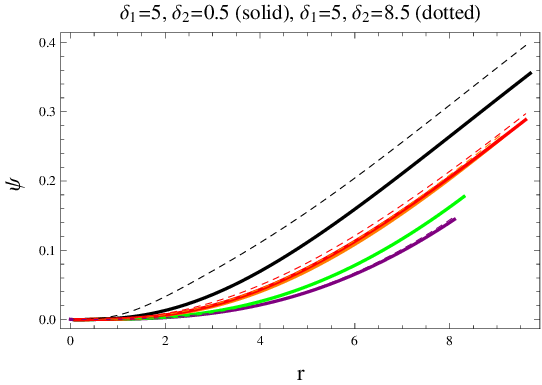,width=0.35\linewidth}
\epsfig{file=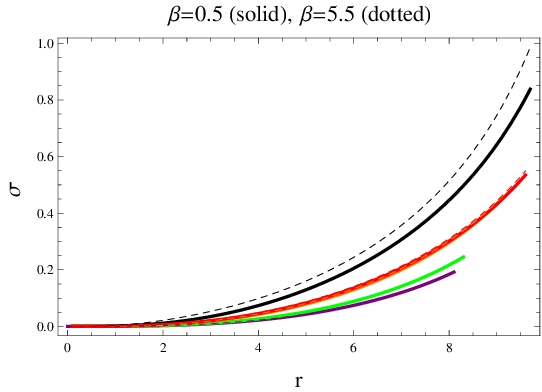,width=0.35\linewidth}\epsfig{file=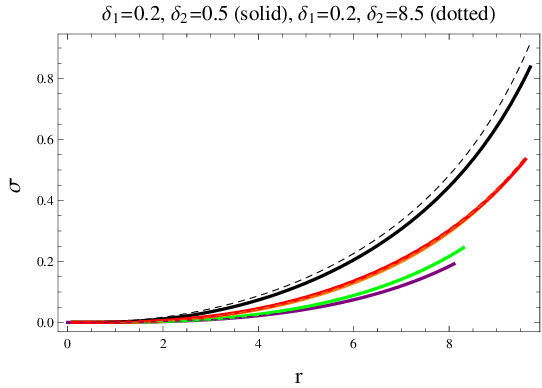,width=0.35\linewidth}\epsfig{file=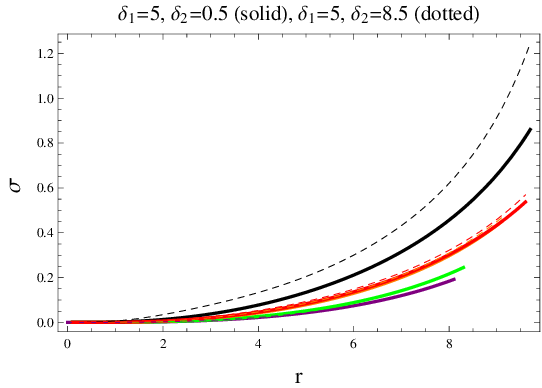,width=0.35\linewidth}
\caption{Physical factors for model 1 (first row) and model 2
(second and third rows).}
\end{figure}

\section{Conclusions}

We have explored the possible existence of spherically symmetric
anisotropic quark configurations for two different models of
$f(\mathcal{G})$ gravity. The field equations have been formulated
corresponding to both the models, characterizing a static spherical
fluid distribution. We have then introduced bag constant in the
gravitational equations through the use of MIT bag model to explore
the quarks' interior. As candidates of quark strange stars, we have
taken Her X-I, LMC X-4, 4U 1820-30, PSR J 1614-2230, and Vela X-I
into account along with their observational data from the literature
(Table \textbf{1}). The equations of motion have been linked with
the MIT model to make sure the presence of the bag constant in them.
Further, we have worked out the modified equations under the use of
Tolman IV metric potentials to make their unique solution possible.
It is then observed that this ansatz involves a triplet of unknowns
which have been determined through the implementation of junction
conditions (Table \textbf{2}). Tables \textbf{3} and \textbf{4}
present the numerical values of $\mathfrak{B}$ for both models whose
expression has been obtained through null radial pressure at the
interface. Additionally, different values of model parameters and
bag constant \eqref{23} have been chosen to graphically explore the
resulting solutions. For instance,
\begin{itemize}
\item For model 1, we have adopted $\beta=0.5,~5.5$.
\item For model 2, we have chosen $\delta_{1}=0.2,~5$ and
$\delta_{2}=0.5,~8.5$.
\end{itemize}

The fluid triplet has been plotted and observed to be consistent
with their required behavior such that they are maximum and finite
at $r=0$ and reach their minimum at the spherical junction. It has
been noticed that the considered structures become more
dense/massive for model 1. Also, a remarkable observation has been
made about the values of bag constant which lie within their
acceptable range. All energy conditions have been satisfied,
guaranteeing the viability of our both models. Certain tests have
been applied on the developed models to check their physical
stability. Finally, we have concluded that the modified Gauss-Bonnet
gravity under its first considered functional form yields physically
relevant results for both parametric values. However, for the second
functional form, the strange stars 4U 1820-30, Vela X-1 and PSR J
1614-2230 are unstable for $\delta_{1}=0.2$ and $\delta_{2}=8.5$.

Similar to our present manuscript, Biswas et al. \cite{63} also
found an acceptable range for the bag constant in their study of
strange stars, which is consistent with our findings. Naseer and
Sharif \cite{60} used the MIT bag EoS to analyze the impact on
several quark stars in the context of non-minimally coupled gravity,
and they determined that the star 4U 1820-30 satisfies all physical
requirements, in contrast with our current results. Das and Debnath
\cite{61} discussed the essential characteristics and physical
viability of two strange stars using the respective EoS, proposing
them as examples of acceptable behavior. Our analysis offers a more
comprehensive study of the internal properties of strange stars
compared to the work done in \cite{62}. Sharif and Hassan \cite{64}
have chosen a specific model of modified Gauss-Bonnet gravity to
examine quark stars and found that the star PSR J1614-2230 becomes
unstable for higher parameter values. It must be stated here that
the results of this theory can be reduced in GR for $\beta=0$ and
$\delta_1=0=\delta_2$ under models 1 and 2, respectively.

\section*{Appendix A}
\renewcommand{\theequation}{A\arabic{equation}} \setcounter{equation}{0}
The derivatives of the Gauss-Bonnet term up to second order are
\begin{align}\nonumber
\mathcal{G}'&=\frac{1}{2 r^3e^{-2
\lambda}}\big[{\nu''} \big(r \big(3 e^{\lambda}-7\big) {\lambda'}+4
\big(e^{\lambda}-1\big)\big)-2 r \nu ^{'''}
\big(e^{\lambda}-1\big)+{\nu'}^2 \\\nonumber&\times
\big(r{\lambda'}\big(e^{\lambda}-2\big) +2
\big(e^{\lambda}-1\big)\big)+{\nu'} \big(r
\big(\big(e^{\lambda}-3\big) {\lambda''}-2 \big(e^{\lambda}-1\big)
{\nu''}\big)\\\label{10b}&+r \big(6-e^{\lambda}\big) {\lambda'}^2-2
\big(e^{\lambda}-3\big) {\lambda'}\big)\big]^{-1},
\\\nonumber \mathcal{G}''&=-\frac{1}{2r^4 e^{-2
\lambda}}\big[2 {\nu''} \big\{2 r^2 \big(e^{\lambda}-5\big)
{\lambda'}^2-r^2 \big(2 e^{\lambda}-5\big) {\lambda''}+2 r \big(3
e^{\lambda}-7\big) {\lambda'}\\\nonumber&+6
\big(e^{\lambda}-1\big)\big\}+2 r^2 \big(e^{\lambda}-1\big)
{\nu''}^2+{\nu'}^2 \big\{r^2 \big(2-e^{\lambda}\big) {\lambda''}+r^2
\big(e^{\lambda}-4\big) {\lambda'}^2\\\nonumber&+4 r
\big(e^{\lambda}-2\big) {\lambda'}+6
\big(e^{\lambda}-1\big)\big\}-{\nu'} \big\{{\lambda'} \big(4 r^2
\big(e^{\lambda}-2\big) {\nu''}-3 \big(r^2 \big(e^{\lambda}-6\big)
{\lambda''}\\\nonumber&-2 e^{\lambda}+6\big)\big)+r^2
\big(e^{\lambda}-12\big) {\lambda'}^3-r \big(r \big(2 \nu ^{'''}
\big(e^{\lambda}-1\big)-\big(e^{\lambda}-3\big) \nu
^{'''}\big)\\\nonumber&-8 \big(e^{\lambda}-1\big) {\nu''}+4
\big(e^{\lambda}-3\big) {\lambda''}\big)+4 r \big(e^{\lambda}-6\big)
{\lambda'}^2\big\}-r \big\{\nu ^{'''} \big( \big(5
e^{\lambda}-11\big)
\\\label{10c}&\times r{\lambda'}+8 \big(e^{\lambda}-1\big)\big)-2 r
\nu ^{''''} \big(e^{\lambda}-1\big)\big\}\big]^{-1}.
\end{align}

\noindent\\ Equations \eqref{8}-\eqref{10} take the form after
inserting the values of $\mathcal{G}$ and its derivatives as
\begin{align}\nonumber
\rho&=\frac{1}{8 \pi r^6 e^{-4 \lambda}}\big[16 \beta  r^2 \big(3
e^{2 \lambda}-35 e^{\lambda}+42\big) \lambda '^3 \nu
'+\big(e^{\lambda}-1\big) \big\{64 \beta  \big(r^2 \big(2
e^{\lambda}-5\big) \lambda ''\\\nonumber&-6 e^{\lambda}+6\big) \nu
''+16 \beta \nu '^2 \big(2 r^2 \big(e^{\lambda}-2\big) \lambda
''+\big(e^{\lambda}-1\big) \big(r^2 \nu''-12\big)\big)-48 \beta
r^2\\\nonumber&\times \big(e^{\lambda}-1\big) \nu ''^2+4 \beta  r^2
\big(e^{\lambda}-1\big) \nu '^4+r \big(r \big(r^2 e^{3 \lambda}-64
\beta \big(e^{\lambda}-1\big) \nu ^{''''}\big)+256
\beta\\\nonumber&\times \big(e^{\lambda}-1\big) \nu ^{'''}\big)+32
\beta  r \nu ' \big(r \big(\big(e^{\lambda}-3\big) \lambda ^{'''}-2
\big(e^{\lambda}-1\big) \nu ^{'''}\big)-4 \big(e^{\lambda}-3\big)
\lambda ''\\\nonumber&+8 \big(e^{\lambda}-1\big) \nu
''\big)\big\}+\lambda ' \big\{r \big(r \big(r^3 e^{3 \lambda}+64
\beta \big(7-10 e^{\lambda}+3 e^{2 \lambda}\big) \nu ^{'''}\big)-64
\beta \big(7 e^{2 \lambda}\\\nonumber&-24 e^{\lambda}+17\big) \nu
''\big)-16 \beta \nu ' \big(r^2 \big(45-48 e^{\lambda}+7 e^{2
\lambda}\big) \lambda ''-\big(e^{\lambda}-1\big) \big(r^2 \big(9
e^{\lambda}\\\nonumber&-19\big) \nu ''+12
\big(e^{\lambda}-3\big)\big)\big)-8 \beta  r^2 \big(3-4
e^{\lambda}+e^{2 \lambda}\big) \nu '^3-32 \beta  r \big(11-16
e^{\lambda}+5\\\nonumber&\times e^{2 \lambda}\big) \nu '^2\big\}-4
\beta r \lambda '^2 \big\{4 r \big(61-64 e^{\lambda}+11 e^{2
\lambda}\big) \nu ''+r \big(11 e^{2 \lambda}-54 e^{\lambda}+47\big)
\nu '^2\\\label{11a}&-8 \big(33-34 e^{\lambda }+5 e^{2 \lambda
}\big) \nu '\big\}\big]^{-1},\\\nonumber P_{r}&=\frac{1}{8 \pi r^5
e^{-4 \lambda}} \big[\nu ' \big\{r \big(r^3 e^{3 \lambda}+32 \beta
\big(e^{2 \lambda }-4 e^{\lambda}+3\big) \nu ^{'''}\big)-32 \beta
\big(e^{\lambda}-3\big) \big(r \big(e^{\lambda}\\\nonumber&-3\big)
\lambda '+2 \big(e^{\lambda }-1\big)\big) \nu ''\big\}-r
\big(e^{\lambda}-1\big) \big\{r^2 e^{3 \lambda}+16 \beta
\big(e^{\lambda}-1\big) \nu ''^2\big\}-
\big(e^{\lambda}-3\big)\\\nonumber&\times 8 \beta\big\{r
\big(e^{\lambda}-3\big) \lambda '+4 \big(e^{\lambda }-1\big)\big\}
\nu '^3+4 \beta  \nu '^2 \big\{4 r \big(\big(3-e^{\lambda}\big)^2
\lambda ''+\big(e^{2 \lambda}-6 e^{\lambda }\\\label{11b}&+5\big)
\nu ''\big)+8 \big(e^{\lambda }-3\big)^2 \lambda '+3 r \big(e^{2
\lambda}-10 e^{\lambda}+21\big) \lambda '^2\big\}-4 \beta r
\big(e^{\lambda}-1\big)^2 \nu '^4\big]^{-1},\\\nonumber
P_{\bot}&=\frac{1}{32 \pi r^5 e^{-4 \lambda}}\big[r \nu '^2 \big\{r
\big(r^3 e^{3 \lambda}-64 \beta \big(e^{\lambda}-3\big) \lambda
^{'''}+64 \beta \big(e^{\lambda}-1\big) \nu ^{'''}\big)-32 \beta (3
r\\\nonumber&-8) \big(e^{\lambda}-3\big) \lambda ''-64 \beta
\big(e^{\lambda}-1\big) \big(e^{\lambda}-3 r+3\big) \nu ''\big\}-2 r
\nu ''\big\{r \big(r^3 e^{3 \lambda}+64 \beta
\big(e^{\lambda}\\\nonumber&-1\big) \nu ^{'''}\big)+32 \beta
\big(e^{2 \lambda}-6 e^{\lambda}+5\big) \nu ''\big\}-32 \beta
\big(e^{\lambda}-1\big) \nu '^3 \big(r^2 \lambda ''+2 r^2 \nu ''+6
r\\\nonumber&-12\big)+2 \nu ' \big\{r^2 \big(r^2 e^{3 \lambda}+64
\beta \big(e^{\lambda}-1\big) \nu ^{''''}\big)+32 r \beta (3 r-8)
\big(e^{\lambda}-1\big) \nu ^{'''} -32\\\nonumber&\times \beta
\big(r^2 \big(3 e^{\lambda}-7\big) \lambda ''+6 (r-2)
\big(e^{\lambda}-1\big)\big) \nu ''\big\}-64 \beta r^2
\big(e^{\lambda}-12\big) \lambda '^3 \nu '^2+\lambda
'\\\nonumber&\times \big\{-r \nu '\big(r \big(r^3 e^{3 \lambda}+64
\beta \big(5 e^{\lambda}-11\big) \nu ^{'''}\big)-32 \beta \big((12-9
r) e^{\lambda}+2 e^{2 \lambda}+21 r\\\nonumber&-38\big) \nu
''\big)+32 \beta \nu '^2 \big(6 r^2 \big(e^{\lambda}-6\big) \lambda
''-r^2 \big(3 e^{\lambda}-5\big) \nu ''+6
(r-2)\big(e^{\lambda}-3\big)\big)\\\nonumber&-2 r^2 \big(r^2 e^{3
\lambda}-32 \beta \big(3 e^{\lambda}-7\big) \nu ''^2\big)+32 \beta
r^2 \big(e^{\lambda}-2\big) \nu '^4-32 \beta r \nu '^3\big(7-e^{2
\lambda}\\\nonumber&+(3 r-2) e^{\lambda}-6 r\big) \big\}+16 \beta r
\lambda '^2 \nu ' \big\{4 r \big(3 e^{\lambda }-14\big) \nu ''+2 r
\big(e^{\lambda}-2\big) \nu '^2+\nu
'\big(e^{\lambda}\\\label{11c}&\times(6r-10) -e^{2 \lambda}-36
r+87\big) \big\}-16 \beta r \big(e^{2 \lambda}-6 e^{\lambda}+5\big)
\nu '^4\big]^{-1}.
\end{align}

\section*{Appendix B}
\renewcommand{\theequation}{B\arabic{equation}} \setcounter{equation}{0}
The matter triplet \eqref{11a}-\eqref{11c} in terms of Tolman IV
potentials are given by
\begin{align}\nonumber
\rho&=\frac{1}{8 \pi \mathbb{C}^8 \big(\mathbb{A}^2+2
r^2\big)^6}\big[\mathbb{A}^4 \mathbb{C}^4 \big(3 \mathbb{A}^6
\mathbb{C}^4+4416 \mathbb{A}^4 \beta +22656 \mathbb{A}^2 \beta
\mathbb{C}^2\\\nonumber&+\mathbb{A}^8 \mathbb{C}^2 \big(32 \pi  B
\mathbb{C}^2-3\big)+25920 \beta  \mathbb{C}^4\big)+32 r^{12} \big(64
\pi  B \mathbb{C}^8-216 \beta \\\nonumber&-9 \mathbb{C}^6\big)+48
r^{10} \big\{\mathbb{A}^2 \big(-288 \beta +128 \pi  B
\mathbb{C}^8-17 \mathbb{C}^6\big)+2 \mathbb{C}^2 \big(144 \beta
+\mathbb{C}^6\big)\big\}\\\nonumber&+48 r^8 \big\{5 \mathbb{A}^2
\big(\mathbb{C}^8+48 \beta  \mathbb{C}^2\big)+4 \mathbb{A}^4
\big(-88 \beta +40 \pi  B \mathbb{C}^8-5 \mathbb{C}^6\big)-272 \beta
\mathbb{C}^4\big\}\\\nonumber&+8 r^6 \big\{6 \mathbb{A}^4 \big(5
\mathbb{C}^8+928 \beta  \mathbb{C}^2\big)+4224 \mathbb{A}^2 \beta
\mathbb{C}^4+\mathbb{A}^6 \big(-384 \beta +640 \pi  B
\mathbb{C}^8\\\nonumber&-75 \mathbb{C}^6\big)+384 \beta
\mathbb{C}^6\big\}+6 r^4 \big\{4 \mathbb{A}^6 \big(5
\mathbb{C}^8+1568 \beta \mathbb{C}^2\big)-160 \mathbb{A}^4 \beta
\mathbb{C}^4\\\nonumber&-8576 \mathbb{A}^2 \beta
\mathbb{C}^6+\mathbb{A}^8 \big(1024 \beta +320 \pi  B
\mathbb{C}^8-35 \mathbb{C}^6\big)\big\}+3 \mathbb{A}^2 \mathbb{C}^2
r^2 \big\{2 \mathbb{A}^6 \big(5 \mathbb{C}^6\\\nonumber&-1792 \beta
\big)-19584 \mathbb{A}^4 \beta  \mathbb{C}^2-19072 \mathbb{A}^2
\beta \mathbb{C}^4+\mathbb{A}^8 \mathbb{C}^4 \big(128 \pi  B
\mathbb{C}^2-13\big)\\\label{11d}&+4608 \beta
\mathbb{C}^6\big\}\big]^{-1},\\\nonumber P_{r}&=-\frac{1}{24 \pi
\mathbb{C}^8 \big(\mathbb{A}^2+2 r^2\big)^7}\big[-\mathbb{A}^6
\mathbb{C}^2 \big\{3 \mathbb{A}^6 \big(1024 \beta
+\mathbb{C}^6\big)+16704 \mathbb{A}^4 \beta
\mathbb{C}^2\\\nonumber&+28032 \mathbb{A}^2 \beta
\mathbb{C}^4+\mathbb{A}^8 \big(3 \mathbb{C}^4-32 \pi  B
\mathbb{C}^6\big)+14400 \beta \mathbb{C}^6\big\}+64 r^{14} \big(-72
\beta \\\nonumber&+64 \pi B \mathbb{C}^8-3 \mathbb{C}^6\big)+64
r^{12} \big\{\mathbb{A}^2 \big(-228 \beta +224 \pi  B
\mathbb{C}^8-11 \mathbb{C}^6\big)-\mathbb{C}^8\\\nonumber&+48 \beta
\mathbb{C}^2\big\}+16 r^{10} \big\{-16 \mathbb{A}^2
\big(\mathbb{C}^8-44 \beta \mathbb{C}^2\big)+\mathbb{A}^4 \big(-1120
\beta +1344 \pi B \mathbb{C}^8\\\nonumber&-71 \mathbb{C}^6\big)-160
\beta \mathbb{C}^4\big\}+16 r^8 \big\{-\mathbb{A}^4 \big(25
\mathbb{C}^8+528 \beta \mathbb{C}^2\big)-1680 \mathbb{A}^2 \beta
\mathbb{C}^4\\\nonumber&+\mathbb{A}^6 \big(-1056 \beta +1120 \pi  B
\mathbb{C}^8-65 \mathbb{C}^6\big)+384 \beta \mathbb{C}^6\big\}-4
\mathbb{A}^2 r^6 \big\{16 \mathbb{A}^4 \big(5
\mathbb{C}^8\\\nonumber&+752 \beta \mathbb{C}^2\big)+18080
\mathbb{A}^2 \beta \mathbb{C}^4+\mathbb{A}^6 \big(4352 \beta -2240
\pi  B \mathbb{C}^8+145 \mathbb{C}^6\big)\\\nonumber&+2688 \beta
\mathbb{C}^6\big\}+4 \mathbb{A}^2 r^4 \big\{-5 \mathbb{A}^6 \big(7
\mathbb{C}^8-64 \beta \mathbb{C}^2\big)+10544 \mathbb{A}^4 \beta
\mathbb{C}^4\\\nonumber&+25728 \mathbb{A}^2 \beta
\mathbb{C}^6+\mathbb{A}^8 \big(-1024 \beta +672 \pi  B
\mathbb{C}^8-49 \mathbb{C}^6\big)+16128 \beta
\mathbb{C}^8\big\}\\\nonumber&+\mathbb{A}^4 r^2 \big\{-32
\mathbb{A}^6 \big(\mathbb{C}^8-944 \beta \mathbb{C}^2\big)+78592
\mathbb{A}^4 \beta  \mathbb{C}^4+100224 \mathbb{A}^2 \beta
\mathbb{C}^6\\\label{11e}&+\mathbb{A}^8 \big(4096 \beta +448 \pi  B
\mathbb{C}^8-37 \mathbb{C}^6\big)+54144 \beta
\mathbb{C}^8\big\}\big]^{-1},
\\\nonumber P_{\bot}&=-\frac{1}{8
\pi \mathbb{C}^8 \big(\mathbb{A}^2+r^2\big) \big(\mathbb{A}^2+2
r^2\big)^7}\big[\mathbb{A}^8 \mathbb{C}^4
\big(\mathbb{A}^2+\mathbb{C}^2\big) \big(\mathbb{A}^6
\mathbb{C}^2+576 \mathbb{A}^2 \beta\\\nonumber& +576 \beta
\mathbb{C}^2\big)+64 r^{16} \big(72 \beta +5 \mathbb{C}^6\big)+1536
\beta  r^{15} \big(\mathbb{A}^2+2 \mathbb{C}^2\big)+64 r^{14}
\big\{\mathbb{A}^2 \\\nonumber&\times\big(252 \beta +17
\mathbb{C}^6\big)-3 \big(\mathbb{C}^8+48 \beta
\mathbb{C}^2\big)\big\}+768 \beta  r^{13} \big(5 \mathbb{A}^4+6
\mathbb{A}^2 \mathbb{C}^2-8 \mathbb{C}^4\big)\\\nonumber&+16 r^{12}
\big\{\mathbb{A}^4 \big(1904 \beta +103 \mathbb{C}^6\big)-32
\mathbb{A}^2 \big(\mathbb{C}^8+26 \beta \mathbb{C}^2\big)+800 \beta
\mathbb{C}^4\big\}+768 \beta  r^{11}\\\nonumber&\times \big(8
\mathbb{A}^6+15 \mathbb{A}^4 \mathbb{C}^2+4 \mathbb{C}^6\big)+16
r^{10} \big\{4 \mathbb{A}^6 \big(344 \beta +23
\mathbb{C}^6\big)-\mathbb{A}^4 \big(33 \mathbb{C}^8\\\nonumber&+4336
\beta \mathbb{C}^2\big)-3280 \mathbb{A}^2 \beta  \mathbb{C}^4-384
\beta \mathbb{C}^6\big\}+384 \mathbb{A}^2 \beta  r^9 \big(22
\mathbb{A}^6+57 \mathbb{A}^4 \mathbb{C}^2\\\nonumber&-52
\mathbb{C}^6\big)+4 r^8 \big\{\mathbb{A}^8 \big(215
\mathbb{C}^6-1536 \beta \big)-12 \mathbb{A}^6 \big(5
\mathbb{C}^8+2448 \beta \mathbb{C}^2\big)\\\nonumber&-7200
\mathbb{A}^4 \beta \mathbb{C}^4+30336 \mathbb{A}^2 \beta
\mathbb{C}^6\big\}+768 \mathbb{A}^2 \beta r^7 \big(8 \mathbb{A}^8+12
\mathbb{A}^6 \mathbb{C}^2-45 \mathbb{A}^4
\mathbb{C}^4\\\nonumber&-65 \mathbb{A}^2 \mathbb{C}^6+18
\mathbb{C}^8\big)+4 \mathbb{A}^2 r^6 \big\{\mathbb{A}^8 \big(85
\mathbb{C}^6-2048 \beta \big)-5 \mathbb{A}^6 \big(\mathbb{C}^8+1408
\beta \mathbb{C}^2\big)\\\nonumber&+37552 \mathbb{A}^4 \beta
\mathbb{C}^4+52608 \mathbb{A}^2 \beta \mathbb{C}^6-16128 \beta
\mathbb{C}^8\big\}+384 \mathbb{A}^4 \beta r^5 \big(4 \mathbb{A}^8-15
\mathbb{A}^6 \mathbb{C}^2\\\nonumber&-96 \mathbb{A}^4
\mathbb{C}^4-55 \mathbb{A}^2 \mathbb{C}^6+90
\mathbb{C}^8\big)+\mathbb{A}^4 r^4 \big\{\mathbb{A}^8 \big(1024
\beta +89 \mathbb{C}^6\big)+8 \mathbb{A}^6 \big(3
\mathbb{C}^8\\\nonumber&+2816 \beta \mathbb{C}^2\big)+98368
\mathbb{A}^4 \beta \mathbb{C}^4+20352 \mathbb{A}^2 \beta
\mathbb{C}^6-132480 \beta \mathbb{C}^8\big\}-768 \mathbb{A}^6 \beta
\mathbb{C}^2 r^3\\\nonumber&\times \big(4 \mathbb{A}^6+9
\mathbb{A}^4 \mathbb{C}^2-16 \mathbb{A}^2 \mathbb{C}^4-36
\mathbb{C}^6\big)+\mathbb{A}^6 \mathbb{C}^2 r^2 \big\{14
\mathbb{A}^8 \mathbb{C}^4+3 \mathbb{A}^6 \big(3
\mathbb{C}^6\\\nonumber&-512 \beta \big)-14784 \mathbb{A}^4 \beta
\mathbb{C}^2-61440 \mathbb{A}^2 \beta \mathbb{C}^4-67392 \beta
\mathbb{C}^6\big\}+384 \mathbb{A}^8 \beta \mathbb{C}^4 r \big(4
\mathbb{A}^4\\\label{11f}&+17 \mathbb{A}^2 \mathbb{C}^2+18
\mathbb{C}^4\big)\big]^{-1}.
\end{align}

\noindent\\ The anisotropic factor for model 1 is given by
\begin{align}\nonumber
\Delta&=\frac{1}{12 \pi  \mathbb{C}^8 \big(\mathbb{A}^2+r^2\big)
\big(\mathbb{A}^2+2 r^2\big)^7}\big[\mathbb{A}^8 \mathbb{C}^2
\big\{-3 \mathbb{A}^6 \big(512 \beta +\mathbb{C}^6\big)-9216
\mathbb{A}^4 \beta \mathbb{C}^2\\\nonumber&-15744 \mathbb{A}^2 \beta
\mathbb{C}^4+\mathbb{A}^8 \mathbb{C}^4 \big(16 \pi  \mathfrak{B}
\mathbb{C}^2-3\big)-8064 \beta \mathbb{C}^6\big\}+64 r^{16}
\big(-144 \beta \\\nonumber&+32 \pi \mathfrak{B} \mathbb{C}^8-9
\mathbb{C}^6\big)+32 r^{14} \big\{\mathbb{A}^2 \big(-1056 \beta +288
\pi  \mathfrak{B} \mathbb{C}^8-65 \mathbb{C}^6\big)+8
\big(\mathbb{C}^8\\\nonumber&+60 \beta \mathbb{C}^2\big)\big\}+32
r^{12} \big\{\mathbb{A}^2 \big(19 \mathbb{C}^8+848 \beta
\mathbb{C}^2\big)+2 \mathbb{A}^4 \big(-968 \beta +280 \pi
\mathfrak{B} \mathbb{C}^8\\\nonumber&-53 \mathbb{C}^6\big)-640 \beta
\mathbb{C}^4\big\}+16 r^{10} \big\{\mathbb{A}^4 \big(29
\mathbb{C}^8+6592 \beta \mathbb{C}^2\big)+4000 \mathbb{A}^2 \beta
\mathbb{C}^4\\\nonumber&+2 \mathbb{A}^6 \big(-1576 \beta +616 \pi
\mathfrak{B} \mathbb{C}^8-103 \mathbb{C}^6\big)+768 \beta
\mathbb{C}^6\big\}+4 r^8 \big\{36992 \mathbb{A}^6 \beta
\mathbb{C}^2\\\nonumber&-1600 \mathbb{A}^4 \beta \mathbb{C}^4-46080
\mathbb{A}^2 \beta \mathbb{C}^6+\mathbb{A}^8 \big(-1984 \beta +3360
\pi \mathfrak{B} \mathbb{C}^8-525
\mathbb{C}^6\big)\big\}\\\nonumber&+2 \mathbb{A}^2 r^6
\big\{\mathbb{A}^6 \big(9408 \beta  \mathbb{C}^2-100
\mathbb{C}^8\big)-120192 \mathbb{A}^4 \beta  \mathbb{C}^4-134784
\mathbb{A}^2 \beta \mathbb{C}^6\\\nonumber&+\mathbb{A}^8 \big(768
\beta +2912 \pi  \mathfrak{B} \mathbb{C}^8-449
\mathbb{C}^6\big)+64512 \beta \mathbb{C}^8\big\}+2 \mathbb{A}^4 r^4
\big\{-\mathbb{A}^6 \big(61 \mathbb{C}^8\\\nonumber&+9024 \beta
\mathbb{C}^2\big)-43584 \mathbb{A}^4 \beta \mathbb{C}^4+35520
\mathbb{A}^2 \beta \mathbb{C}^6+\mathbb{A}^8 \big(-768 \beta +784
\pi  \mathfrak{B} \mathbb{C}^8\\\nonumber&-125
\mathbb{C}^6\big)+129024 \beta \mathbb{C}^8\big\}+\mathbb{A}^6 r^2
\big\{-31 \mathbb{A}^6 \big(\mathbb{C}^8-512 \beta
\mathbb{C}^2\big)+53120 \mathbb{A}^4 \beta
\mathbb{C}^4\\\nonumber&+128256 \mathbb{A}^2 \beta
\mathbb{C}^6+\mathbb{A}^8 \big(2048 \beta +240 \pi  \mathfrak{B}
\mathbb{C}^8-41 \mathbb{C}^6\big)+120960 \beta
\mathbb{C}^8\big\}\\\nonumber&-2304 \beta  r^{15}
\big(\mathbb{A}^2+2 \mathbb{C}^2\big)-1152 \beta  r^{13} \big(5
\mathbb{A}^4+6 \mathbb{A}^2 \mathbb{C}^2-8 \mathbb{C}^4\big)-1152
\beta r^{11}\\\nonumber&\times \big(8 \mathbb{A}^6+15 \mathbb{A}^4
\mathbb{C}^2+4 \mathbb{C}^6\big)-576 \mathbb{A}^2 \beta  r^9 \big(22
\mathbb{A}^6+57 \mathbb{A}^4 \mathbb{C}^2-52
\mathbb{C}^6\big)-1152\\\nonumber&\times \mathbb{A}^2 \beta  r^7
\big(8 \mathbb{A}^8+12 \mathbb{A}^6 \mathbb{C}^2-45 \mathbb{A}^4
\mathbb{C}^4-65 \mathbb{A}^2 \mathbb{C}^6+18 \mathbb{C}^8\big)-576
\mathbb{A}^4 \beta  r^5\big(4 \mathbb{A}^8 \\\nonumber&-15
\mathbb{A}^6 \mathbb{C}^2-96 \mathbb{A}^4 \mathbb{C}^4-55
\mathbb{A}^2 \mathbb{C}^6+90 \mathbb{C}^8\big)+1152 \mathbb{A}^6
\beta \mathbb{C}^2 r^3\big(4 \mathbb{A}^6-16 \mathbb{A}^2
\mathbb{C}^4\\\label{16a}&+9 \mathbb{A}^4 \mathbb{C}^2-36
\mathbb{C}^6\big)-576 \mathbb{A}^8 \beta \mathbb{C}^4 r \big(4
\mathbb{A}^4+17 \mathbb{A}^2 \mathbb{C}^2+18
\mathbb{C}^4\big)\big]^{-1}.
\end{align}\\
\textbf{Data Availability Statement:} This manuscript has no
associated data.

\end{document}